\documentclass[aip,jcp,reprint]{revtex4-1}
\usepackage{amsmath}
\usepackage{amssymb}
\usepackage{amsfonts}

\usepackage{fancybox}
\usepackage{times}
\usepackage{latexsym}
\usepackage{pifont}
\usepackage{graphicx}
\usepackage{pstricks}

\newcommand{\intensity}[2]{$#1\times10^{#2}\mbox{ W/cm}^{2}$}

\newcommand{\bv}[1]{\mbox{\boldmath $#1$}}

\def\jpbo{{\em J Phys B: At Mol Phys\/}}        
\def\jpcm{{\em J Phys: Cond Matt\/}}   

\def\pra{{\em Phys Rev A\/}}
\def\prb{{\em Phys Rev B\/}}
\def\pre{{\em Phys Rev E\/}}

\def\prl{{\em Phys Rev Lett\/}}

\def\jcp{{\em J Chem Phys\/}}        

\newcommand{\gradi}[1]{\tilde{\bv{\nabla}}_{\!#1}}
\newcommand{\lape}[1]{\nabla_{#1}^2}

\begin{document}

\title{Multielectron effects in high harmonic generation in N$_{\mbox{\scriptsize\bfseries\sffamily 2}}$ and
       benzene: simulation using a non-adiabatic quantum molecular dynamics 
       approach for laser-molecule interactions}

\author{Daniel Dundas}

\affiliation{Atomistic Simulation Centre, School of Mathematics and Physics, \\
             Queen's University Belfast, \\
             Belfast BT7 1NN, Northern Ireland, UK.} 
\date{\today}

\begin{abstract}
A mixed quantum-classical approach is introduced which allows the dynamical response of 
molecules driven far from equilibrium to be modeled. This method is applied to the 
interaction of molecules with intense, short-duration laser pulses. The electronic response 
of the molecule is described using time-dependent density functional theory (TDDFT) and the 
resulting Kohn-Sham equations are solved numerically using finite difference techniques in 
conjunction with local and global adaptations of an underlying grid in curvilinear 
coordinates. Using this approach, simulations can be carried out for a wide range of 
molecules and both all-electron and pseudopotential calculations are possible. The approach 
is applied to the study of high harmonic generation in N$_2$ and benzene using 
linearly-polarized laser pulses and, to the best of our knowledge, the results for benzene 
represent the first TDDFT calculations of high harmonic generation in benzene using linearly 
polarized laser pulses. For N$_2$ an enhancement of the cut-off harmonics is observed whenever 
the laser polarization is aligned perpendicular to the molecular axis. This enhancement is 
attributed to the symmetry properties of the Kohn-Sham orbital that responds predominantly 
to the pulse. In benzene we predict that a suppression in the cut-off harmonics occurs 
whenever the laser polarization is aligned parallel to the molecular plane. We attribute 
this suppression to the symmetry-induced response of the highest-occupied molecular orbital. 
\end{abstract}

\maketitle


\begin{section}{Introduction}
Molecules driven out of equilibrium by external forces are of fundamental
importance in many areas of science and technology. In such a non-equilibrium 
situation, a non-adiabatic coupling exists between electrons and ions which can 
induce charge and energy transfer across the molecule on a femtosecond 
timescale~\cite{zewail:1994}. These charge and energy transfer processes are of 
extreme importance in the design of electronic devices~\cite{joachim:2000}, 
probes and sensors~\cite{willard:2003}, and in the areas of condensed matter 
and plasma physics, medicine and biochemistry. 

Of particular interest and importance is the interaction of molecules with 
ultra-short laser pulses. In this case, the laser couples to the electrons in 
the molecule and this coupling can lead to ionization and subsequent 
dissociation of the molecule if the laser intensity is high enough. Indeed, 
even for low laser intensities current flow in molecular electronic devices can
be both induced and suppressed~\cite{hanggi:2007,franco:2007,li:2007}. Over the 
last decade much effort has focused on the possibility of using ultra-short
laser pulses to control chemical reactions. Experimental studies have already
used genetic algorithms to create optimal pulse profiles in order to break 
specific bonds in complex 
molecules~\cite{brixner:2003,wollenhaupt:2007,laarmann:2008}. In these 
experiments Ti:sapphire laser pulses (wavelength, $\lambda\sim$ 800nm) were 
employed with the interaction taking place over a timescale of 10s of femtoseconds. 
Alternatively, few-cycle optical pulses have been used to steer dissociation by 
varying the carrier-envelope phase of the 
pulse~\cite{kling:2006,kremer:2009,ullrich:2010b}. More recently, the use of 
attosecond pulses has been considered for controlling the electron dynamics 
directly. These pulses generally operate at high laser frequencies (VUV to 
x-ray wavelengths) and can be created either through high harmonic generation (HHG) in 
gases~\cite{baltuska:2003,hentschel:2001,paul:2001,drescher:2001} or in 
free-electron laser sources. Much of the work on attosecond pulses has 
focused on controlling processes such as electron 
localization~\cite{becker:2008,sansone:2010,singh:2010} as this control is seen 
as the the crucial first step in controlling chemical reactions: a fuller
review of this subject is given by Krausz and Ivanov~\cite{krauz:2009}.

The study of multielectron effects in the response of molecules to intense laser pulses has attracted
much attention in recent years. This interest stems from the complexity of the electronic structure 
in molecules together with the experimental ability to control the orientation between the molecules and 
the laser pulse using a variety of alignment techniques.  While initial studies considered multielectron 
effects in ionization~\cite{talebpour:1996,guo:1998,muth:2000,dewitt:2001,chu:2004,dundas:2004b,faisal:2006,%
ullrich:2010,chu:2011}, recent studies have considered the role of multielectron effects in  
HHG~\cite{hay:2000,zdanska:2003,baer:2003,ceccherini:2001,denalda:2004,mcfarland:2008,kajumba:2008,le:2009,%
lee:2010,telnov:2009,kamta:2009,smirnova1:2009,smirnova2:2009,smirnova3:2009,mairesse:2010,vozzi:2010,%
heslar:2011,bandrauk:2010,bandrauk:2011}. 
Studying HHG as a function of alignment provides us not only with an insight to the role of
molecular electronic structure in HHG, but also with a technique for manipulating the strength of the
high harmonics emitted. Indeed, the sub-femtosecond timing information encoded in these high harmonics 
can be used as a tool to dynamically image the molecular orbitals
themselves~\cite{itatani:2004,marangos:2008} and, if several orbitals contribute to the HHG process,
sub-femtosecond electron dynamics in the molecule can be directly imaged~\cite{smirnova1:2009}. Therefore, 
an understanding of multielectron effects in molecular HHG is crucial to the development of these imaging
techniques.

Any model for describing the interaction of molecules with ultra-short laser 
pulses must be able to handle length, time and energy scales that can span 
several orders of magnitude. In the first instance we need to model processes 
occurring on timescales ranging from the sub-femtosecond for 
electronic motion to 100s of femtoseconds for the dissociation of the molecule. 
Secondly, the laser wavelength can vary from optical to VUV wavelengths which
alters how the laser couples to the electrons in the molecule: while optical 
pulses couple predominantly with the valence electrons in the molecule, 
attosecond pulses can couple directly to the innermost electrons. Lastly, in
describing processes such as ionization and HHG we must be able
to accurately describe the emission of high-energy electrons
over length scales of several hundred Bohr radii. For simple molecular systems, 
such as one- and two-electron diatomic molecules, solution of the time-dependent
Schr\"{o}dinger equation (TDSE) is possible~\cite{becker:2008,chelk:1992,kono:1997,%
lein:2003,dundas:2003,bandrauk:1998,lein:2003,bandrauk:2004,saenz:2010}. However, 
describing more complex molecules requires further approximation. These
approximations range from simple models which handle only crucial aspects of 
the dynamics to full ab initio simulations of the electrons and ions in the 
presence of the laser pulse. 

One ab initio method that is 
widely used to treat electronic dynamics is time-dependent 
density functional theory (TDDFT)~\cite{runge:1984}. The TDDFT method has been
applied extensively to the study of molecules and clusters subject to external 
forces~\cite{saalmann:1996,saalmann:1998,kunert:2003,uhlmann:2003,dundas:2004,%
calvayrac:2000,castro:2004}. In many cases this application involves coupling 
the quantum treatment of the electronic degrees of freedom to a classical treatment 
of the nuclear degrees of freedom, resulting in a method known as the non-adiabatic 
quantum molecular dynamics (NAQMD) method~\cite{saalmann:1996,saalmann:1998,kunert:2003}. 
Implementations of NAQMD have generally used basis-set~\cite{saalmann:1996,saalmann:1998,%
kunert:2003,uhlmann:2003} or grid~\cite{dundas:2004,calvayrac:2000,castro:2004} techniques. 
While grid approaches offer better opportunities for code parallelization, allowing 
efficient scaling to large problem sizes, they do suffer from the drawback of requiring 
small grid spacings in order to deal with moving ions on the grid. This problem can be 
overcome by using grid adaptation techniques employing both local and global 
adaptations~\cite{kono:1997,gygi:1992,devenyi:1994,hamann:1995,fattal:1996,perez:1998,%
gygi:1995,perez:1995,modine:1997,castro:2006,dundas:2000}. 

In this paper we set out our NAQMD approach on a real space grid that allows the 
use of both local and global adaptations. The novelty of the approach lies in the 
range of methods that are implemented and that, in implementing these methods together, 
a wide range of problems can be studied. The approach can handle either all
of the electrons or only a subset, depending on the laser-pulse used, and its 
implementation opens up the possibility of carrying out ab initio simulations
of the response of complex molecules to intense laser pulses of arbitrary
polarization. This is in contrast to many other implementations of TDDFT which have been 
optimised to efficiently study ionization and HHG in either diatoms irradiated by
linearly polarized light~\cite{telnov:2009,heslar:2011} 
or in polyatomic molecules, like benzene, irradiated by circularly polarized light~\cite{baer:2003}.
The paper is arranged as follows. In Sec.~\ref{sec:naqmd} a 
set of equations of motion for a system of quantum mechanical electrons and 
classical ions is derived using a Lagrangian formalism. This derivation allows 
for velocity-dependent terms to be introduced in cases where either a moving basis 
or moving grid is introduced. In Sec.~\ref{sec:tddft} the equations of motion are 
re-expressed whenever TDDFT is used to describe the electronic system. 
A number of local and global coordinate transformations are described in 
Sec.~\ref{sec:adaptive} which will allow a standard finite difference grid to be warped 
in different regions of space. Sec.~\ref{sec:calculation} describes how the NAQMD approach 
is implemented in a real-space code. In Sec.~\ref{sec:results} the method is applied to study 
the influence of multielectron effects in HHG in molecules. In particular, our calculations for HHG in 
benzene predict a suppression of the high-order harmonics as the alignment between the molecular plane and the
laser polarization direction varies. Some conclusions are drawn in Sec.~\ref{sec:conclusions}.

Unless otherwise stated, atomic units are used throughout.
\end{section}


\begin{section}{The non-adiabatic quantum molecular dynamics approach}
\label{sec:naqmd}
We consider a system consisting of $N_e$ quantum-mechanical electrons 
and $N_n$ classical ions. 
The electrons are described by their many-body wavefunction 
$\Psi(\bv{r}_e, t)$, where $\bv{r}_e = \left\{\bv{r}_1, \dots, \bv{r}_{N_e}\right\}$
denotes the electron position vectors (for the time being we will ignore spin).
The ions are described by their 
trajectories $\bv{R} = \left\{\bv{R}_1(t), \dots, \bv{R}_{N_n}(t)\right\}$ 
and momenta $\bv{P} = \left\{\bv{P}_1(t), \dots, \bv{P}_{N_n}(t)\right\}$. 
For ion $k$, we denote its mass and charge by $M_k$ and $Z_k$ respectively.

In order to derive a set of equations of motion for the electrons and ions we
will use a Lagrangian
formalism~\cite{todorov:2001,niehaus:2005}. We start from the Lagrangian
\begin{eqnarray}
\label{eq:lagrangian}
{\cal L} & = & 
\mbox{i}\int d\bv{r}_e\Psi^\star(\bv{r}_e, t)\dot{\Psi}(\bv{r}_e, t) \nonumber\\
& - &
\int d\bv{r}_e\Psi^\star(\bv{r}_e, t) H(\bv{r}_e, \bv{R}, t)\Psi(\bv{r}_e, t) \nonumber\\
&+ & 
\frac{1}{2}\sum_{k = 1}^{N_n}M_k \dot{\bv{R}}_k^2(t) - V_{nn}(\bv{R}),
\end{eqnarray}
where
\begin{eqnarray}
V_{nn}(\bv{R}) = \sum_{k<k^\prime}^{N_n} \frac{Z_kZ_{k^\prime}}{\left|\bv{R}_k -
\bv{R}_{k^\prime}\right|},
\end{eqnarray}
denotes the Coulomb repulsion between the ions
and $H(\bv{r}_e, \bv{R}, t)$ is the time-dependent Hamiltonian for the electrons which depends 
parametrically on the ion coordinates. The Hamiltonian can be written as
\begin{equation}
H(\bv{r}_e, \bv{R}, t) = 
\sum_{i=1}^{N_e}
\left[-\frac{1}{2}\lape{i} + V_{\mbox{\scriptsize ext}}(\bv{r}_i, \bv{R}, t)\right] + 
V_{ee}(\bv{r}_e),
\label{eqn:hamil}
\end{equation}
where $\lape{i}$ is the Laplacian with respect to the
coordinates of electron $i$. In this Hamiltonian 
\begin{equation}
V_{ee}(\bv{r}_e) = \sum_{i<j}^{N_e} \frac{1}{\left|\bv{r}_i - \bv{r}_j\right|},
\end{equation}
is the Coulomb repulsion between electrons and
\begin{equation}
V_{\mbox{\scriptsize ext}}(\bv{r}_i, \bv{R}, t) = 
V_{\mbox{\scriptsize ions}}(\bv{r}_i, \bv{R}, t) + 
U_{\mbox{\scriptsize elec}}(\bv{r}_i, t),
\label{eq:pot_ext}
\end{equation}
is the external potential consisting of $U_{\mbox{\scriptsize elec}}(\bv{r}_i, t)$, the 
interaction between electron $i$ and the applied laser field, and 
\begin{eqnarray}
V_{\mbox{\scriptsize ions}}(\bv{r}_i, \bv{R}) & = & \sum_{k=1}^{N_n} V_{\mbox{\scriptsize
ion}}(\bv{r}_i, \bv{R}_k) \nonumber\\
& = & 
-\sum_{k=1}^{N_n}\frac{Z_k}{\left|\bv{r}_i - \bv{R}_k\right|},\label{eq:ion_pot}
\end{eqnarray}
the Coulomb interaction between electron $i$ and all ions. Additionally, in Eq.~(\ref{eq:lagrangian})
\begin{equation}
\int d\bv{r}_e = \int d\bv{r}_1\cdots\int d\bv{r}_{N_e},
\end{equation}
refers to integration over all electron coordinates.

The equations of motion for the electrons and ions can be obtained by considering variations of the
wavefunction and ion trajectories that leave the action, $\cal A$, stationary i.e.
\begin{equation}
\delta {\cal A} = 
\delta \int_{t_0}^{t_1} {\cal L}dt = 
0.
\end{equation}
This results in the Euler-Lagrange equations of motion
\begin{eqnarray}
\frac{\partial{\cal L}}{\partial \Psi^\star} & = &  \frac{d}{dt}\left({\frac{\partial{\cal L}}{\partial
\dot{\Psi}^\star}}\right) ,\label{eq:lagpsistar}\\[0.2cm]
\frac{\partial{\cal L}}{\partial \Psi} & = &  \frac{d}{dt}\left({\frac{\partial{\cal L}}{\partial
\dot{\Psi}}}\right) ,\label{eq:lagpsi}\\[0.2cm]
\frac{\partial{\cal L}}{\partial \bv{R}_k} & = &  \frac{d}{dt}\left({\frac{\partial{\cal L}}{\partial
\dot{\bv{R}}_k}}\right).\label{eq:lagrk}
\end{eqnarray}
Eq.~(\ref{eq:lagpsistar}) leads to the TDSE
\begin{equation}
\label{eq:tdse_eom}
\mbox{i}\frac{\partial}{\partial t} \Psi(\bv{r}_e, t) =
H(\bv{r}_e, \bv{R}, t)\Psi(\bv{r}_e, t) ,
\end{equation}
while Eq.~(\ref{eq:lagpsi}) gives its complex conjugate.
Eq.~(\ref{eq:lagrk}) leads to the equation of motion for the ions
\begin{eqnarray}
M_k \ddot{\bv{R}}_k = &-& \int d\bv{r}_e\Psi^\star(\bv{r}_e,t)
                         \left(\gradi{k}H(\bv{r}_e, \bv{R},
		    t)\right)\Psi(\bv{r}_e,t) \nonumber\\[0.2cm]
		    &-& \gradi{k}V_{nn}(\bv{R}),\label{eq:hfgrid}
\end{eqnarray}
where $\gradi{k}$ denotes the gradient operator with respect to the ionic coordinates of 
ion $k$.

The Lagrangian formalism has the benefit of allowing the equations of
motion for electrons and ions to be easily derived in situations where either a finite basis set is used 
or where dynamical locally-adaptive grids -- in which the grid is adapted in a small region around 
the instantaneous ionic
positions -- are used. For instance, when an incomplete basis set of atom-centred Gaussian functions 
is used, the  resulting equations of motion have been derived by several 
authors~\cite{saalmann:1996,todorov:2001,niehaus:2005}. In that case, velocity-dependent   
forces (the so-called Pulay forces~\cite{pulay:1969}) are introduced. The use of dynamical,
locally-adaptive grids will also introduce Pulay-type forces on the ions. This point will be returned to in
Sec.~\ref{sec:adaptive}.

It is important to note that a mixed quantum-classical description of electron-ion dynamics has a number of
drawbacks. For example, a classical description of heavy ions can be justified when considering
processes which occur over a timescale of several femtoseconds. However, when considering processes that involve 
light ions (such as hydrogen atoms) quantum effects can become important, in which case a classical description will 
no longer be appropriate. This is particularly true when considering the response of simple diatomic molecules, such
as H$_2^+$ and H$_2$, to intense laser pulses. For these systems, quantum treatments of electrons and 
ions can be carried out and these clearly show the importance of quantum nuclear 
motion~\cite{lein:2002,dundas:2003,kreibach:2004,martin:2005,bandrauk:2008}. For instance, Bandrauk and co-workers
clearly show that a a quantum description of nuclear motion is important for HHG in H$_2^+$~\cite{bandrauk:2008}.

\end{section}


\begin{section}{Time-dependent density functional treatment of the electron
                dynamics}
\label{sec:tddft}
In Sec.~\ref{sec:naqmd} we derived a set of equations of motion for a system of
quantum mechanical ions and classical ions. The TDSE derived for the many-body electronic
wavefunction can only be solved in full dimensionality for the simplest few body systems. In this section
we will consider a TDDFT description of electronic structure~\cite{runge:1984}. 
 
In an earlier paper
a set of equations of motion was derived for the electrons and ions using a static
real-space grid~\cite{dundas:2004}. Here, we
derive the equations of motion, using the Lagrangian formalism, for the Kohn-Sham orbitals in TDDFT that will
allow us to consider the more general case of a
grid that can deforms around the instantaneous ion positions. 
In the TDDFT method, the total $N_e$-electron Kohn-Sham wavefunction is written 
as a single determinant of electron orbitals $\psi_{i\sigma}(\bv{r}, t)$ with the electron density given by  
\begin{equation}
\label{eq:density}
n(\bv{r}, t) = \sum_{\sigma = \downarrow,\uparrow} n_{\sigma}(\bv{r}, t) = \sum_{\sigma = \downarrow,\uparrow}\sum_{i=1}^{N_\sigma} \left|\psi_{i\sigma}(\bv{r}, t)\right|^2,
\end{equation}
where $N_{\sigma}$ is the number of Kohn-Sham orbitals for spin state $\sigma$ and $N_e = N_\downarrow +
N_\uparrow$.

In order to derive the equations of motion we consider the Lagrangian
\begin{eqnarray}
\label{eq:lagrangian_tdse}
{\cal L} & = & 
\mbox{i}\sum_{\sigma = \downarrow,\uparrow} \sum_{i=1}^{N_\sigma}\int d\bv{r}
\psi_{i\sigma}^\star(\bv{r}, t)\dot{\psi}_{i\sigma}(\bv{r}, t) \nonumber\\
& + &
\frac{1}{2}\sum_{\sigma = \downarrow,\uparrow} \sum_{i=1}^{N_\sigma}
\int d\bv{r}\psi_{i\sigma}^\star(\bv{r}, t)
\lape{} \psi_{i\sigma}(\bv{r}, t) \nonumber\\
& - &
\int d\bv{r}n(\bv{r}, t) \left(V_{\mbox{\scriptsize ext}}(\bv{r}, \bv{R}, t)  + 
\frac{1}{2}\int d\bv{r}^\prime \frac{n(\bv{r}^\prime, t)}{\left|\bv{r} -
\bv{r}^\prime\right|}\right) \nonumber\\
&-&
A_{\mbox{\scriptsize xc}}[n_\downarrow, n_\uparrow] \nonumber\\
&+ & 
\frac{1}{2}\sum_{k = 1}^{N_n}M_k \dot{\bv{R}}_k^2(t) - V_{nn}(\bv{R}).
\end{eqnarray}
In this equation, $A_{\mbox{\scriptsize xc}}[n_\downarrow, n_\uparrow]$ is the exchange-correlation
action functional of TDDFT. The equations of motion for the Kohn-Sham orbitals and ions can be 
obtained in a similar fashion to Eqs.~(\ref{eq:lagpsistar})--(\ref{eq:lagrk}) with the many-body 
wavefunction, $\Psi^\star$, in Eq.~(\ref{eq:lagpsistar}) replaced by the Kohn-Sham orbitals, 
$\psi_{i\sigma}^\star(\bv{r}, t)$. From these Euler-Lagrange equations we obtain the time-dependent 
Kohn-Sham equations
\begin{equation}
\mbox{i}\frac{\partial}{\partial t} \psi_{i\sigma}(\bv{r}, t) =
H_{\mbox{\scriptsize ks},\sigma}(\bv{r}, \bv{R}, t)\psi_{i\sigma}(\bv{r}, t).
\label{eq:tdks_eom}
\end{equation}
In this equation 
\begin{equation}
\label{eq:tdks_ham}
H_{\mbox{\scriptsize ks},\sigma}(\bv{r}, \bv{R}, t) = -\frac{1}{2}\lape{} + V_{\mbox{\scriptsize eff},\sigma}(\bv{r},
\bv{R}, t),
\end{equation}
is the Kohn-Sham Hamiltonian and 
\begin{equation}
\label{eq:tdks_veff}
V_{\mbox{\scriptsize eff},\sigma}(\bv{r}, \bv{R}, t) = V_{\mbox{\scriptsize ext}}(\bv{r}, \bv{R}, t) + 
V_{\mbox{\scriptsize H}}(\bv{r}, t) + V_{\mbox{\scriptsize xc},\sigma}(\bv{r}, t) , 
\end{equation}
where
\begin{equation}
\label{eq:hartree}
V_{\mbox{\scriptsize H}}(\bv{r}, t) = \int d\bv{r}^\prime \frac{n(\bv{r}^\prime, t)}{\left|\bv{r} -
\bv{r}^\prime\right|},
\end{equation}
is the Hartree potential, $V_{\mbox{\scriptsize ext}}(\bv{r}, \bv{R}, t)$ is defined from 
Eq.~(\ref{eq:pot_ext}) and $V_{\mbox{\scriptsize xc}\sigma}(\bv{r}, t)$ is the exchange-correlation 
potential.

The corresponding equations of motion for the ions are obtained from Eq.~(\ref{eq:lagrk}) and are analogous to
those obtained in Eq.~(\ref{eq:hfgrid}), namely
\begin{eqnarray}
\label{eq:hf_tddft}
M_k \ddot{\bv{R}}_k = &-& \sum_{\sigma=\downarrow,\uparrow}\int d\bv{r}
n_\sigma(\bv{r},t)\left(\gradi{k}H_{\mbox{\scriptsize ks},\sigma}(\bv{r}, \bv{R},
		    t)\right) \nonumber\\[0.2cm]
		    &-& \gradi{k}V_{nn}(\bv{R}).
\end{eqnarray}

At this point we have a set of equations of motion for electrons and ions which introduce a non-adiabatic
coupling between the two subsystems. We now show how these equations can be solved
numerically using a real-space grid in adaptive curvilinear coordinates.
\end{section}


\begin{section}{Adaptive curvilinear coordinates}
\label{sec:adaptive}
Adaptive curvilinear coordinates have been widely used in electronic structure calculations, 
both in plane-wave basis-set methods~\cite{gygi:1992,devenyi:1994,hamann:1995,fattal:1996,perez:1998}
and real-space grid 
techniques~\cite{gygi:1995,perez:1995,modine:1997,castro:2006,dundas:2000,kono:1997,kono:1999}. 
In this approach
either a local or global warping of space is carried out and in some situations a combination of both 
is used. In finite
difference methods this warping of space leads to a high density of grid points around the centres of
adaptation while in plane-wave methods the warping gives rise to an effective energy cut-off which varies
locally~\cite{gygi:1993}.
 
In general we will
consider the physical space to be described by Cartesian coordinates and then consider a transformation to a
set of generalised curvilinear coordinates (which may, or may not, be orthogonal) that act as the computational
space. The Kohn-Sham equations will then be solved in these curvilinear coordinates using finite difference techniques
(see Sec.~\ref{sec:finite_diff}). In this scenario, the grid points in curvilinear coordinates are equally spaced which results in 
efficient nearest neighbour communication patterns when the resulting computer code is parallelized (see Sec.~\ref{sec:parallel}).

In the following, we describe a range of global and local coordinate transformations that we have 
implemented. Our underlying curvilinear system will be described by the coordinates 
$\left(\zeta^1, \zeta^2, \zeta^3\right)$.
Global adaptation will transform these to a set of coordinates $\left(u(\zeta^1), v(\zeta^2),
w(\zeta^3)\right)$. A
local adaptation can then be applied to these coordinates to transform to Cartesian coordinates 
$\left(x, y, z\right)$ so that
\begin{equation}
\begin{array}{lclcl}
\zeta^1 &             & u(\zeta^1) &             & x(u, v, w) = x(\zeta^1, \zeta^2, \zeta^3)\\[0.2cm]
\zeta^2 & \Rightarrow & v(\zeta^2) & \Rightarrow & y(u, v, w) = y(\zeta^1, \zeta^2, \zeta^3)\\[0.2cm]
\zeta^3 &             & w(\zeta^3) &             & z(u, v, w) = z(\zeta^1, \zeta^2, \zeta^3)
\end{array}.
\end{equation}
The transformation between Cartesian and curvilinear coordinates is described by the 
Jacobian matrix
\begin{equation}
\bv{J} = \left(\begin{array}{ccc}
                \frac{\displaystyle\partial x^1}{\displaystyle\partial \zeta^1} & 
		\frac{\displaystyle\partial x^1}{\displaystyle\partial \zeta^2} & 
		\frac{\displaystyle\partial x^1}{\displaystyle\partial \zeta^3}\\[0.4cm]
                \frac{\displaystyle\partial x^2}{\displaystyle\partial \zeta^1} & 
		\frac{\displaystyle\partial x^2}{\displaystyle\partial \zeta^2} & 
		\frac{\displaystyle\partial x^2}{\displaystyle\partial \zeta^3}\\[0.4cm]
                \frac{\displaystyle\partial x^3}{\displaystyle\partial \zeta^1} & 
		\frac{\displaystyle\partial x^3}{\displaystyle\partial \zeta^2} & 
		\frac{\displaystyle\partial x^3}{\displaystyle\partial \zeta^3}
               \end{array}
	  \right)
\end{equation}
where each element can be written as $J^i_\alpha$ and its determinant, $|J| = \det \bv{J}$, describes how the volume element changes. 
The metric in Cartesian coordinates, $g_{ij} = \delta_{ij}$, can be written in matrix form
as the identity matrix, i.e. $\bv{g} = \bv{I}$. In curvilinear coordinates the metric tensor 
transforms to  $g_{\alpha\beta}$ whose elements can be written in matrix form as 
$\bv{g} = \bv{J}^T\bv{J}$.

\begin{subsection}{Globally adaptive curvilinear coordinates}
\label{sec:global_adapt}
Globally adaptive coordinates allow one particular curvilinear coordinate to be scaled independently 
of the others. Such scaling techniques have been widely used whenever a particular density of points 
is required along a given axis. For example, in the treatment of linear molecules a global scaling has 
been used in cylindrical coordinates to give a high density of points near the molecular 
axis~\cite{dundas:2000,kono:1997,kono:1999}. There are a number of ways to implement a global scaling. 
We will detail three such techniques which transform the curvilinear coordinate $\zeta$ to the scaled 
coordinate $u$.
\begin{enumerate}
\item Using the transformation 
\begin{equation}
\label{eq:sinh_mesh}
u(\zeta) = \sinh\left(\frac{\zeta}{\alpha}\right),
\end{equation}
where $\alpha$ is a parameter used to control the maximum extent of the grid in $u$, we obtain 
a high density of grid points near the origin and a low density of grid points far from the origin.
\item Under the transformation 
\begin{equation}
u(\zeta) = \zeta\left(\frac{\zeta^n}{\zeta^n + \alpha^n}\right)^\nu,
\end{equation}
where $n$ takes on integer values, $\nu$ takes on half-integer values in general and $\alpha$ is a real number,
we obtain a high density of grid points near origin and an equidistant grid spacing far from the origin~\cite{kono:1999}. 
The parameter $\alpha$ controls where the transition between the regularly-spaced and densely-spaced grid occurs. 
\item Consider the transformation
\begin{equation}
u(\zeta) = \left\{
\begin{array}{cl}
\zeta & |\zeta| \le \zeta_f\\[0.4cm]
\zeta + d_{\max}\left(\frac{\displaystyle\zeta - \zeta_f}{\displaystyle\zeta_f - \zeta_{\max}}\right)^5
& |\zeta|> \zeta_f
\end{array}\right.,
\end{equation}
where $d_{\max} = \zeta_{\max} - u_{\max}$, $\zeta_{\max}$ is the maximum value of the unscaled coordinate, $\zeta_f$ is the point where the flat region ends,
$u_{\max}$ is the maximum value of the scaled coordinate required and where $u(-\zeta) = -u(\zeta)$ when $\zeta< -\zeta_f$. This scaling gives rise to a grid that has an
equidistant spacing near origin and a low density of grid points far from the origin. This is a specific
case of the general global backdrop described by Modine et al~\cite{modine:1997} in which $u$ and its
first four derivatives match at $|\zeta| = \zeta_f$.
\end{enumerate}
When both local and global adaptations are used only the third global transformation 
is employed.
\end{subsection}

\begin{subsection}{Locally adaptive curvilinear coordinates}
\label{sec:local_adapt}
\begin{figure*}
\centerline{\hfill\includegraphics[width=6cm]{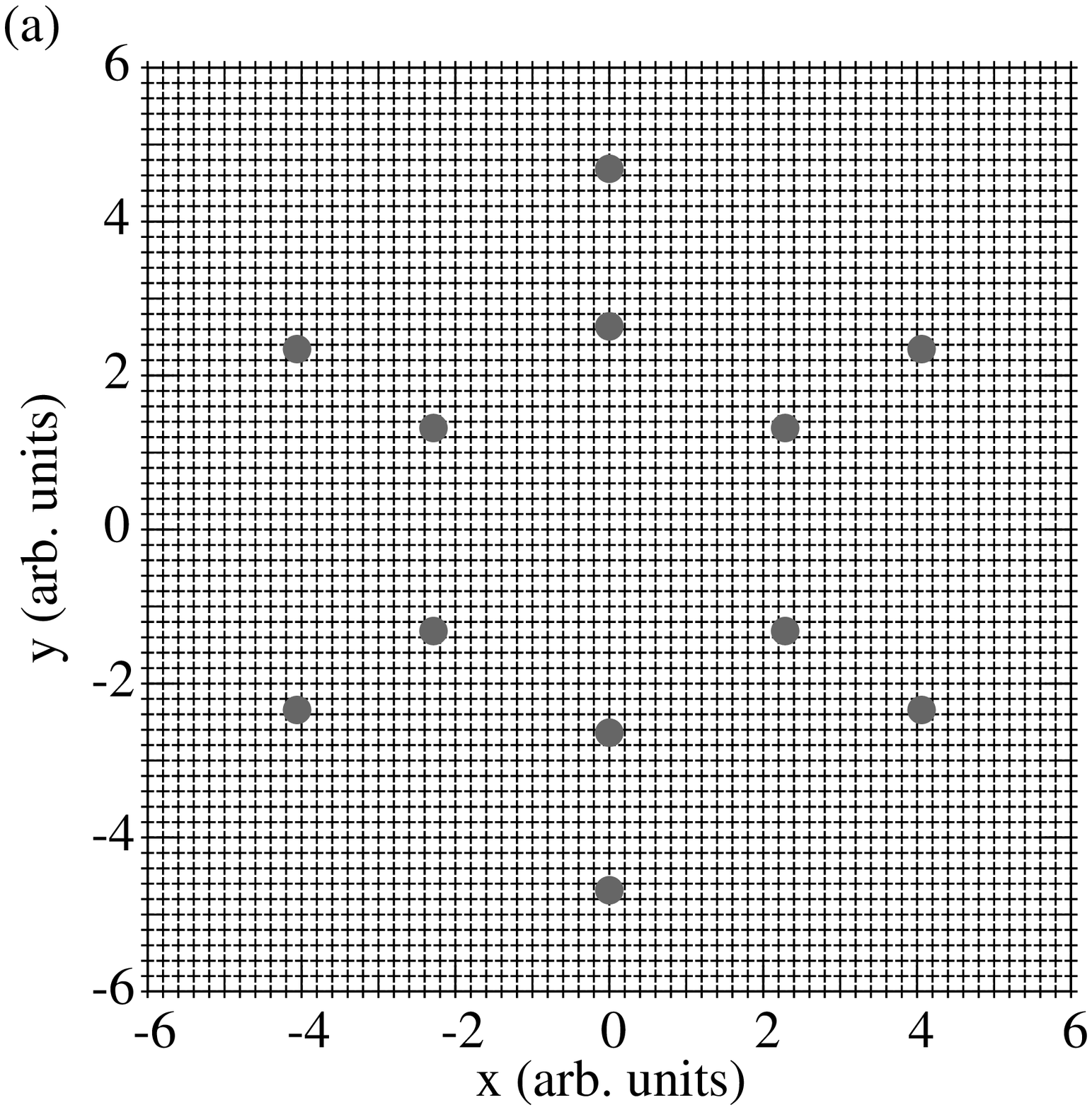}%
            \hfill\includegraphics[width=6cm]{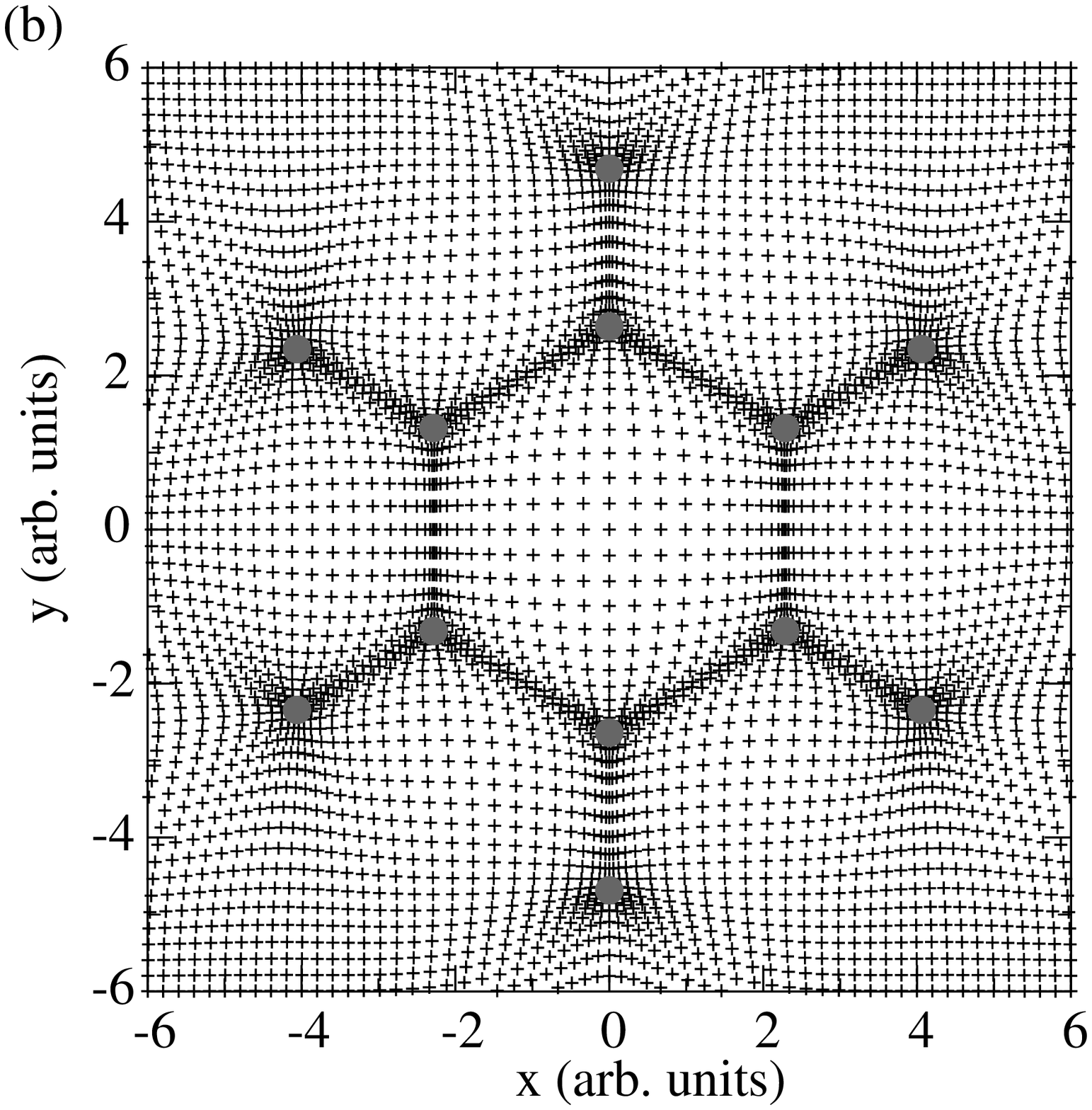}%
	    \hfill}
\caption{Illustration of local adaptation around the 12 atoms in the benzene molecule (denoted by the grey circles) using 
the transformation between Cartesian and curvilinear coordinates described by Eq.~(\protect\ref{eq:adapt_tranform}). 
The molecule lies in the $x-y$ plane and for clarity only two dimensions are shown. (a) shows the Cartesian 
grid under no adaptation, while (b) shows the deformation of the grid when adaptation is present.}
\label{fig:adapt}
\end{figure*}
In finite-difference electronic structure calculations the error is greatest in regions around the ions. 
This error is usually minimized by using a combination of high-order finite difference formulae and small grid 
spacings~\cite{chelikowsky:1994,beck:2000}. An alternative approach is to carry out a local adaptation of the 
curvilinear coordinates around the ions to give a high density of points in these regions. In the 
following, we implement the adaptive coordinate technique used in the ACRES DFT approach of Modine and 
co-workers~\cite{modine:1997} and describe the transformation from 
generalised curvilinear coordinates 
$\bv{\zeta} = (\zeta^1, \zeta^2, \zeta^3)$ to Cartesian coordinates 
$\bv{r} = (x^1, x^2, x^3) = (x, y, z)$. This procedure can be extended to incorporate the global
adaptations introduced in Sec.~\ref{sec:global_adapt}. The local transformation is
\begin{equation}
\label{eq:adapt_tranform}
\bv{r} = \bv{\zeta} - \sum_{k=1}^{N_n} 
\bv{Q}_k\cdot\Bigr(\bv{\zeta} - \bv{{\cal R}}_k(t)\Bigr)
\exp\left[-\left(\frac{\bv{\zeta} - \bv{{\cal R}}_k(t)}{\sigma_k}\right)^2\right].
\end{equation}
In this expression $\bv{Q}_k$ and $\sigma_k$ are parameters controlling the strength and extent of 
adaptation in the vicinity of ion $k$. In general each $\bv{Q}_k$ is a $3\times 3$ matrix, although 
it is sufficient to consider each $\bv{Q}_k$ to be diagonal. We adjust the elements of $\bv{Q}_k$ to 
satisfy $J_\alpha^i(\bv{{R}}_k) = \left|J(\bv{{R}}_k)\right|^{1/3}\delta_\alpha^i$. If the adaptation 
regions associated with different ions overlap, each adaptation centre attempts to draw grid points 
to itself. As a result of this competition, the highest density of points will not be around the 
exact ionic positions. Therefore, we perform the adaptation around \emph{fictitious} positions 
$\bv{{\cal R}}_k(t)$ which are chosen so that on the fully-adapted grid 
$\bv{r}(\bv{{\cal R}}_k) = \bv{R}_k$. Optimal values for the elements of $\bv{Q}_k$ and 
$\bv{{\cal R}}_k$ are found self-consistently using Jacobi iteration. 

The effect of this transformation is illustrated in Fig.~\ref{fig:adapt} where we consider a local adaptation 
of a finite difference grid around the atoms in benzene, denoted by the 
grey circles. In Fig.~\ref{fig:adapt}(a) we present the Cartesian grid with no adaptation using a grid 
spacing of 0.4 a.u. For clarity we only present the two dimensions $x$ and $y$. Under the 
transformation of Eq.~(\ref{eq:adapt_tranform}) the adapted grid is shown in Fig.~\ref{fig:adapt}(b). The effect 
of this transformation is an increase in grid density in the vicinity of the atoms. In this example we use identical 
adaptations around each atom. However, the form of Eq.~(\ref{eq:adapt_tranform}) is such that each atom can have 
different strengths and extents of adaptation.

Obviously under this coordinate transformation, the distribution of points will change as
the ions move. Since the electronic position vector will  depend implicitly on the ion positions, 
this results in the introduction of Pulay corrections to the forces
defined in Eq.~(\ref{eq:hf_tddft}) \cite{pulay:1969,modine:1997}. In
addition, as the ions move the grid deformation will alter and so a grid regeneration will be
required together with an interpolation of the Kohn-Sham orbitals onto this new grid. In 
situations  where the ions do not move significantly, we can choose a grid deformation
which can be kept static throughout a simulation. In that case,  
Pulay force corrections, grid regeneration and orbital interpolation will not be required. Such an approach is similar to that of Hamann~\cite{hamann:2001}.

Another point to note from Fig.~\ref{fig:adapt} is that when the
Gaussian factor in Eq.~(\ref{eq:adapt_tranform}) becomes sufficiently small, i.e. away from the 
atoms, the curvilinear coordinates reduce to the Cartesian coordinates. In these regions of space, the
finite difference grid will become independent of the ion positions and those terms in
Eqs.~(\ref{eq:tdks_eom})~and~(\ref{eq:hf_tddft}) depending on
derivatives of the electronic coordinates with respect to the ionic coordinates will vanish.

\end{subsection}
\begin{subsection}{The Kohn-Sham equations in adaptive curvilinear coordinates}
\label{sec:differential_adapt}
With these coordinate transformations we can rewrite the time-dependent Kohn-Sham equations in terms of 
the curvilinear coordinates, $\bv{\zeta}$. Referring to the Lagrangian in 
Eq.~(\ref{eq:lagrangian_tdse}) and the Kohn-Sham equations of 
Eq.~(\ref{eq:tdks_eom}) we require expressions for the volume element and the Laplacian operator.
Integrals will transform according to 
\begin{equation}
\label{eq:transform_int}
\int d\bv{r} f(\bv{r}) \rightarrow \int |J| d\bv{\zeta} f(\bv{\zeta}),
\end{equation}
while the Laplacian is given by the Lapace-Beltrami operator
\begin{equation}
\label{eq:laplacian_full}
\nabla^2 = \frac{1}{|J|}\frac{\partial}{\partial \zeta^\alpha}
         |J| g^{\alpha\beta}
         \frac{\partial}{\partial \zeta^\beta},
\end{equation}
where $g^{ij}$ is the contravariant metric tensor and Einstein summation notation is assumed.
Transforming the Kohn Sham orbitals according to
\begin{equation}
\label{eq:ks_transform}
\psi_{i\sigma}(\bv{r}, t) = \left|J\right|^{-1/2} \varphi_{i\sigma}(\bv{r}, t),
\end{equation}
results in the Laplacian operator in the Kohn-Sham equations taking the form
\begin{equation}
\label{eq:laplacian_transform}
\nabla^2 = \frac{1}{\sqrt{\left|J\right|}}\frac{\partial}{\partial
\zeta^\alpha}\left|J\right|g^{\alpha\beta}\frac{\partial}{\partial \zeta^\beta}\frac{1}{\sqrt{\left|J\right|}}.
\end{equation}
\end{subsection}
This is important when a finite difference treatment of the Laplacian is used so that the resulting finite difference
equations remain symmetric, thus allowing the use of unitary time propagation schemes for the solution
of the Kohn-Sham equations. In three-dimensional space the Laplacian will have nine terms, three of which involve 
first derivatives in the same variable. These three terms can be rewritten so that they involve second derivatives in the
same variable while still maintaining the symmetry of the finite difference Laplacian. For example, if we 
consider the Laplacian term involving $\zeta^1$ we can write
\begin{widetext}
\begin{equation}
\label{eq:laplacian_diag}
\frac{1}{\sqrt{\left|J\right|}}\frac{\partial}{\partial
\zeta^1}\left|J\right|g^{11}\frac{\partial}{\partial \zeta^1}\frac{1}{\sqrt{\left|J\right|}}\varphi_{i\sigma}(\bv{r}, t) =
\frac{1}{2}\left[g^{11}\frac{\partial^2}{\partial {\zeta^1}^2} + \frac{\partial^2}{\partial {\zeta^1}^2}g^{11}\right]
\varphi_{i\sigma}(\bv{r}, t) + M^{11} \varphi_{i\sigma}(\bv{r}, t),
\end{equation}
\end{widetext}
where
\begin{eqnarray}
M^{11} = \frac{1}{4|J|^2}\Biggr[ & & \left(|J|^\prime\right)^2g^{11} - 2|J|^2{g^{11}}^{\prime\prime}\nonumber\\
 & - & 
2|J||J|^\prime {g^{11}}^\prime - 2 |J||J|^{\prime\prime}g^{11}\Biggr],
\end{eqnarray}
and $f^\prime$ denotes differentiation of $f$ with respect to $\zeta^1$. As well as preserving the symmetry of the
resulting finite difference equations, this symmetrization also reduces communication overheads when implemented  
as a parallel computer code using domain decomposition. We shall discuss this further in Sec.~\ref{sec:parallel}.
\end{section}


\begin{section}{Implementation of the method}
\label{sec:calculation}
 
A typical calculation requires an initial ionic configuration to be chosen. 
Using this configuration  the density is
calculated self-consistently from the Kohn-Sham equations. 
This density and configuration is then used in the solution of equations of
motion for the electrons and ions in the presence of a laser pulse. 
In this section we describe the technical details of our real-space 
implementation of the NAQMD approach in adaptive curvilinear coordinates.


\begin{subsection}{Discretization of the Kohn-Sham equations using finite differences}
\label{sec:finite_diff}

The curvilinear coordinates $\bv{\zeta} = (\zeta^1, \zeta^2, \zeta^3)$ are discretized on a finite difference grid
with constant grid spacings $(\Delta\zeta^1, \Delta\zeta^2, \Delta\zeta^3)$ in each dimension. For coordinate $\zeta^i$, where $i =
(1, 2, 3)$
we choose a set of $N_{\zeta^i}$ points which cover the range $-\zeta^i_{\rm max} \leq \zeta^i \leq
\zeta^i_{\rm max} $ so that the grid spacing is given by 
\begin{equation}
\Delta \zeta^i= \frac{2\zeta^i_{\rm max}}{N_{\zeta^i}-1}, 
\end{equation} 
and the grid points are denoted by 
\begin{equation}
\label{eq:fd_global}
\zeta^i_{\nu} = -\zeta^i_{\rm max} +(\nu-1) \Delta \zeta^i, \,\,\,\,\,\,\,\,\nu = 1, \dots, N_{\zeta^i}. 
\end{equation} 
Derivatives appearing in the Laplacian -- Eqs.~(\ref{eq:laplacian_full}) and~(\ref{eq:laplacian_diag}) -- are approximated
by central difference formulae. The first derivative of a function, $f(\zeta^i)$, at the point $\zeta^i_{\nu}$ can be approximated
by 
\begin{equation}
   f^{\prime}(\zeta^i_{\nu}) = \frac{1}{\Delta \zeta^i} 
   \sum_{\kappa = -N_{\mbox{\scriptsize fd}}}^{N_{\mbox{\scriptsize fd}}}C^{(1)}_\kappa f(\zeta^i_{\nu +\kappa}),
\label{eqn:fdop1}
\end{equation}
while the second derivative is approximated as
\begin{equation}
   f^{\prime\prime}(\zeta^1_{\nu}) = \frac{1}{(\Delta \zeta^i)^2} 
   \sum_{\kappa = -N_{\mbox{\scriptsize fd}}}^{N_{\mbox{\scriptsize fd}}}C^{(2)}_\kappa f(\zeta^i_{\nu +\kappa}).
\label{eqn:fdop2}
\end{equation}
In these two equations $C^{(1)}_\kappa$  and $C^{(2)}_\kappa$ are the finite difference coefficients and the order of the
finite difference formula is $2N_{\mbox{\scriptsize fd}} + 1$. As a compromise between accuracy and sparsity in the
resulting finite-difference equations, we generally use 5-point formulae, but higher-order formulae (up to 13-point) are also implemented.

Integrals of a function, as defined in Eq.~(\ref{eq:transform_int}), are simply approximated as
\begin{equation}
\label{eq:fd_int}
\int d\bv{r} f(\bv{r}) \approx \Delta \zeta^1\Delta \zeta^2\Delta \zeta^3\sum_{\nu_1 = 1}^{N_{\zeta^1}}
\sum_{\nu_2 = 1}^{N_{\zeta^2}} \sum_{\nu_3 = 1}^{N_{\zeta^3}} |J| f(\zeta^1_{\nu_1}, \zeta^2_{\nu_2}, \zeta^3_{\nu_3}).
\end{equation}
\end{subsection}


\begin{subsection}{Parallelization}
\label{sec:parallel}
The parallel solution of the Kohn-Sham equations is achieved by a domain decomposition in which the 
full finite-difference grid is distributed over processors, with each processor storing all Kohn-Sham 
orbitals for that spatial region. Such a decomposition allows for greater scalability of the code compared 
to a parallelisation over Kohn-Sham orbitals. This is particularly important when studying the interaction
of molecules with intense laser pulses where we must use large spatial grids to hold the ionizing wavepackets. As
we will see below, a grid parallelization will result in communication patterns that only involves halo grid 
points. A parallelization over Kohn-Sham orbitals, on the other hand, would suffer from two major drawbacks. 
Firstly, for operations involving different Kohn-Sham orbitals, such as calculation of the density, 
the whole orbital must be transferred. This results in a large communication overhead. Secondly,
a given system will only have a finite, potentially small, number of Kohn-Sham orbitals. This places a limit
on the maximum number of processors that can be used in a given calculation.

A full decomposition of the finite difference grid is carried out in each spatial dimension. 
In a particular calculation we use $N^i_p$ processors in the $\zeta^i$ direction, 
where $i = (1, 2, 3)$, so that the calculation uses $N_p = N^1_pN^2_pN^3_p$ processors in total.
Each processor in the $\zeta^i$ direction is labeled as 
\begin{equation}
P^i_{\rho} = \rho\,\,\,\,\,\,\,\,\rho = 0, \dots, N^i_p.
\end{equation}
On each
processor we have $N^i_{\mbox{\scriptsize loc}}$ local $\zeta^i$-points along the $\zeta^i$ direction so that 
$N_{\zeta^i} = N^i_{\mbox{\scriptsize loc}}N^i_p$. Each local point on processor $P^i_{\rho}$ in the $\zeta^i$ direction can be mapped
to a global point defined in Eq.~(\ref{eq:fd_global}) by
\begin{equation}
   \zeta^i_{\nu} = -\zeta^i_{\rm max} + 
          (N^i_{\mbox{\scriptsize loc}} \times P^i_{\rho} + \mu - 1) \times \Delta \zeta^i,
\end{equation}
where $\mu = 1,\dots, N^i_{\mbox{\scriptsize loc}}$.

The main communications bottleneck using this decomposition takes place when applying the Laplacian operator. A typical
communications pattern for applying the Laplacian is presented in Fig.~\ref{fig:parallel}. For clarity 
we will limit our discussion to 2D and consider the case of using a 5-point approximation of both first and 
second derivatives appearing in
Eqs.~(\ref{eq:laplacian_transform})~and~(\ref{eq:laplacian_diag}). In this figure the circles denote the points held on one
processor while the squares and triangles denote information required from other processors. When orthogonal curvilinear 
coordinates are used the information denoted by the triangles is not required and thus the use of non-orthogonal coordinate
transformations (such as that described in Sec.~\ref{sec:local_adapt}) increases the communication overhead in a
calculation. The benefit of using Eq.~(\ref{eq:laplacian_diag}) for evaluating those terms in the Laplacian involving derivatives
in the same variable is now apparent. With Eq.~(\ref{eq:laplacian_diag}) only two halo points are required for
communication on each processor boundary. Alternatively, if we use the first derivative finite difference expressions directly in
Eq.~(\ref{eq:laplacian_transform}), this would entail extra communication overhead. For example, the half point rule
suggested by Modine et al~\cite{modine:1997} would result in three halo points being required on each processor boundary.

\begin{figure}
\centerline{\includegraphics[width=6cm]{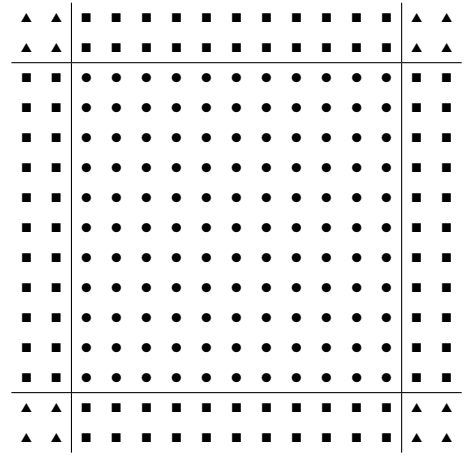}}
\caption{Schematic diagram of communication pattern required for application of the Laplacian given in 
equations~(\protect\ref{eq:laplacian_transform}) and~(\protect\ref{eq:laplacian_diag}) for a 3D domain decomposition of the finite
difference grid. Processor boundaries are represented by the solid lines and for clarity only two dimensions are shown. The grid points local to a given processor are represented by
(\ding{108}) while the halo points required from other processors are denoted by (\ding{110}) and (\ding{115}). Halo points
denoted by (\ding{115}) are required for application of those terms in the Laplacian in 
Eq.~(\protect\ref{eq:laplacian_transform}) that involve derivatives in different variables when using non-orthogonal
curvilinear coordinates. Therefore, the use of non-orthogonal coordinates introduces an increased communications overhead in
a given calculation.}
\label{fig:parallel}
\end{figure}
\end{subsection}


\begin{subsection}{Time propagation}
To propagate the solutions of the NAQMD equations in time we must handle
both electrons, Eq.~(\ref{eq:tdks_eom}), and 
ions, Eq.~(\ref{eq:hf_tddft}). Since the 
ions move more slowly than the electrons, we can use different timesteps for each of these 
subsystems. In the following we denote the time-step for the evolution of the electronic subsystem by 
$\Delta t_e$
while the time-step for the ionic subsystem is denoted by $\Delta t_i$.

\begin{subsubsection}{Time propagation of the Kohn-Sham equations}
Time-propagation of the Kohn-Sham equations requires the use of a unitary
propagator. Given a Kohn-Sham orbital $\varphi_{i\sigma}(\bv{r}, t)$ at a time $t$, the 
orbital at a later time, $t+\Delta t_e$, is obtained by applying the 
unitary time evolution operator $U_{i\sigma}(t+\Delta t_e, t)$, namely
\begin{eqnarray}
\varphi_{i\sigma}(\bv{r}, t + \Delta t_e) & =       & U_{i\sigma}(t +\Delta t_e, t)\varphi_{i\sigma}(\bv{r}, t)\nonumber\\[0.2cm]
                               & \approx & e^{-\mbox{\scriptsize i}H_{\mbox{\scriptsize ks},\sigma}(t)\Delta t_e}\varphi_{i\sigma}(\bv{r}, t). \label{eqn:arnoldi_prop}
\end{eqnarray}
We use the accurate, high-order unitary $n$th-order Arnoldi propagator to
approximate the evolution operator $U_{i\sigma}(t +\Delta t_e, t)$~\cite{arnoldi:1951,smyth:1998,dundas:2004}. 
\end{subsubsection}
\begin{subsubsection}{Time propagation of the ionic equations of motion}

For time-propagation of the classical classical equations of motion we use a velocity Verlet algorithm. In that case we evolve the positions and velocities
of ion $k$ according to 
\begin{eqnarray}
\hspace*{-0.4cm}\bv{R}_k(t + \Delta t_i)       & = & \bv{R}_k(t) + \dot{\bv{R}}_k(t) \Delta t_i + \frac{1}{2}\ddot{\bv{R}}_k(t)(\Delta t_i)^2\nonumber\\[0.2cm]
\hspace*{-0.4cm}\dot{\bv{R}}_k(t + \Delta t_i) & = & \dot{\bv{R}}_k(t) + \frac{\ddot{\bv{R}}_k(t) + \ddot{\bv{R}}_k(t + \Delta t_i)}{2} \Delta t_i,
\end{eqnarray}
where $\ddot{\bv{R}}_k(t)$ is given by Eq.~(\ref{eq:hf_tddft}).
\end{subsubsection}
\end{subsection}


\begin{subsection}{Calculation of the initial state}
\label{sec:initial}
Before time propagation of the NAQMD equations we need to calculate the initial state of the system.
We will consider the system to be in its ground electronic state 
for a given ionic configuration. In that case only the ground-state electronic 
density is required~\cite{runge:1984}. If we
require both the electronic and ionic system to be in equilibrium, this can be achieved by an 
iterative procedure in which the ionic configuration is optimized to minimize the ground-state energy of the system. 
For a given  geometry, the ground-state energy is obtained from solving the time-independent Kohn-Sham
equations self-consistently. The limited memory BFGS method~\cite{liu:1989} is then used to find a better
approximation to the equilibrium geometry. This procedure is repeated until convergence to a desired tolerance
is achieved.

The major bottleneck in this procedure is the self-consistent solution of the Kohn-Sham equations
at each step of the optimization procedure. The method for self-consistent solution of the time-independent Kohn-Sham equations 
is set out in the following six steps.

\begin{enumerate}
\item Get initial guess for Kohn-Sham orbitals, $\varphi_{i\sigma}(\bv{r})$\\[-0.6cm]
\item Calculate charge density $n(\bv{r})$\\[-0.6cm]
\item Calculate effective potential $V_{\mbox{\scriptsize eff},\sigma}(\bv{r})$\\[-0.6cm]
\item Solve Kohn-Sham equations
$$
\left[\frac{1}{2}\nabla^2 + V_{\mbox{\scriptsize eff},\sigma}(\bv{r})\right]\varphi_{i\sigma}(\bv{r})
= \epsilon_i\varphi_{i\sigma}(\bv{r}), \hspace*{0.5cm} i = 1, \dots 
$$
\item Calculate new charge density, $\eta(\bv{r})$ \\[-0.6cm]
\item If  $\left|\eta(\bv{r}) - n(\bv{r})\right| < \varepsilon$ \\[-0.4cm]
 \begin{itemize}
 \item[] Stop\\[-0.6cm]
 \end{itemize}
 else\\[-0.6cm]
 \begin{itemize}
 \item[] Mix densities\\[-0.4cm]
 \item[] Calculate the new effective potential $V_{\mbox{\scriptsize eff},\sigma}(\bv{r})$\\[-0.4cm]
 \item[] Go to  step 4\\[-0.6cm]
 \end{itemize}
end
\end{enumerate}

In the above scheme, $\varepsilon$ denotes the tolerance at which we decide the calculation has converged.
While the convergence criteria outlined in this scheme involves the density, we have also implemented a
convergence test based on the effective potential. We find that for sufficiently small values of 
$\varepsilon$, the results agree well and the computational time is similar.
In addition, the density obtained through the solution of the Kohn-Sham equations is not used directly for the
next iteration. Instead we carry out density mixing using either Anderson~\cite{anderson:1964} or 
Pulay~\cite{pulay:1980} mixing. 

The most time-consuming part of the self-consistent calculation is obtaining the Kohn-Sham eigenpairs at step~4. We
require methods for calculating the ground-state density that are as efficient as possible and allow for effective
parallelization. Several methods are now described.

\begin{subsubsection}{Propagation in imaginary time}
The Kohn-Sham eigenenergies and eigenvectors can be obtained through propagating 
the time-dependent Kohn-Sham equations in imaginary time, subject to the constraint that the 
orbitals maintain orthogonality. This is achieved by making the replacement 
$\Delta \tau = \mbox{i}\Delta t_e$ in  Eq.~(\ref{eqn:arnoldi_prop}). This has the effect of turning 
the Kohn-Sham equations into a set of diffusion equations, in which the
eigenvectors decay at rates proportional to their energies.
Such an approach has been used by several groups~\cite{dundas:2004,hernandez:2007}. 

\end{subsubsection}
\begin{subsubsection}{Thick-restarted Lanczos (TRLan) method}

Lanczos~\cite{lanczos:1950} showed that the eigen-decomposition of the Krylov-subspace Hamiltonian can be used as a first 
step in an iterative scheme to calculate the eigenvalues of the actual Hamiltonian. This 
approach can prove more attractive than time-propagation in imaginary time, especially when
many eigenpairs are required. However, in this case, the Lanczos method becomes computationally expensive due to the 
size of the Krylov subspace required and the cost of maintaining orthogonality of the subspace vectors. A number of 
schemes have been proposed to reduce this cost ranging from partially reorthogonalized schemes~\cite{bekas:2005} which aim to reduce the 
computational cost associated with maintaining orthogonality of the Krylov vectors to restarted methods~\cite{wu:1999} which aim to 
reduce the dimension of the Krylov subspace. 

We have implemented the Thick Restart Lanczos method in the TRLan library~\cite{wu:1999} in order to calculate the eigenenergies and
eigenvectors of the required Kohn-Sham orbitals.
\end{subsubsection}
\begin{subsubsection}{The Chebyshev filtered subspace iteration (CheFSI) method}
As stated earlier, the major bottleneck in the calculation of the ground-state is
the calculation of the Kohn-Sham eigenpairs in each self-consistent cycle. Recently, Zhou et al~\cite{zhou:2006,zhou:2006b} developed an
approach that emphasises the importance of eigenspaces rather than eigenpairs in the self-consistent solution
of the Kohn-Sham equations. The basis of their approach is the fact that the density can be calculated from diagonal of
the density matrix
\begin{equation}
\Pi = \Phi \Phi^\dagger,
\end{equation}
where $\Phi$ is the matrix whose columns are the occupied Kohn-Sham orbitals. For any unitary matrix,
$U$, of appropriate dimension we can write
\begin{equation}
\Pi = (\Phi U) (\Phi U)^\dagger,
\end{equation}
and thus explicit eigenvectors are not required in order to calculate the density, instead any orthonormal
basis of the eigenspace corresponding to the occupied Kohn-Sham states will do. In their approach, the
eigenproblem is solved once in order to provide an approximate eigenspace. This eigenspace must only be
slightly larger than the number of states required. The approximate eigenspace is then filtered in
each cycle of the self-consistency loop using a Chebyshev filter, $p_m(x)$, of order $m$ constructed from the
Hamiltonian, $H_{\mbox{\scriptsize ks},\sigma}$. The application of this filter on the eigenspace, 
$p_m(H_{\mbox{\scriptsize ks},\sigma})\Phi$, should then provide a better approximation to the eigenspace of 
occupied states.

The filter is based upon the fast-growth property of Chebyshev polynomials of the first kind outside 
the interval $[-1, 1]$. It is constructed by obtaining estimates for the lower and upper bound of the 
unwanted part
of the spectrum of $H_{\mbox{\scriptsize ks},\sigma}$. These bounds can easily be calculated: see Zhou et al~\cite{zhou:2006} for details. The
filter is then constructed so that the unwanted part of the spectrum is mapped onto the region 
$[-1, 1]$. In that case, the self-consistency loop becomes
\begin{enumerate}
\item Get initial guess for Kohn-Sham orbitals, $\varphi_{i\sigma}(\bv{r})$\\[-0.6cm]
\item Calculate charge density $n(\bv{r})$\\[-0.6cm]
\item Calculate the effective potential 
$V_{\mbox{\scriptsize eff}, \sigma}(\bv{r})$\\[-0.6cm]
\item Solve Kohn-Sham equations
\vspace*{-0.2cm}
$$
\left[\frac{1}{2}\nabla^2 + V_{\mbox{\scriptsize eff}, \sigma}(\bv{r})\right]\varphi_{i\sigma}(\bv{r})
= \epsilon_i\varphi_{i\sigma}(\bv{r}), \hspace*{0.5cm} i = 1, \dots
$$
\item Calculate new charge density, $\eta(\bv{r})$\\[-0.6cm]
\item If  $\left|\eta(\bv{r}) - n(\bv{r})\right| < \mbox{tol}$ \\[-0.4cm]
 \begin{itemize}
 \item[] Stop\\[-0.6cm]
 \end{itemize}
 else\\[-0.6cm]
 \begin{itemize}
 \item[] Mix densities\\[-0.4cm]
 \item[] Calculate the new effective potential $V_{\mbox{\scriptsize eff}, \sigma}(\bv{r})$\\[-0.4cm]
 \item[] Perform Chebyshev subspace iteration\\[-0.4cm]
 \item[] Go to step 5\\[-0.6cm]
 \end{itemize}
end
\end{enumerate}

At step~4 any appropriate eigensolver can be used to provide an initial approximation to the eigenspace: we use the TRLan 
package.
\end{subsubsection}
\end{subsection}


\begin{subsection}{Calculation of the Hartree potential}
\label{sec:hartree}
The Hartree potential integral in Eq.~(\ref{eq:hartree}) is evaluated by 
solving the corresponding Poisson equation
\begin{equation}
\nabla^2 V_H(\bv{r}, t) = -4\pi n(\bv{r}, t).
\end{equation}
Writing $V_H(\bv{r}, t) = |J|^{-1/2}W_H(\bv{r}, t)$, as in Eq.~(\ref{eq:ks_transform}), results in the 
Laplacian in the Poisson equation having the form given by Eq.~(\ref{eq:laplacian_transform}). 
Thus
\begin{equation}
\frac{1}{\sqrt{\left|J\right|}}\frac{\partial}{\partial
\zeta^\alpha}\left|J\right|g^{\alpha\beta}\frac{\partial}{\partial \zeta^\beta}\frac{1}{\sqrt{\left|J\right|}}W_H(\bv{r}, t) = 
-4\pi |J|^{1/2}n(r, t).
\label{eq:hartree-poisson}
\end{equation}
This equation is solved using a conjugate gradient method on our finite
difference grid. In order to 
apply the correct boundary conditions we consider a multipole expansion of the Hartree
potential on the boundary of the grid so that
\begin{equation}
n(\bv{r})|_{\mbox{\scriptsize boundary}} = \sum_{l=0}^{\infty} \sum_{m = -l}^{l} \frac{4\pi}{2l+1} \frac{1}{r^{l+1}}
Y_{lm}(\hat{\bv{r}}) Q_{lm},
\end{equation}
where 
\begin{equation}
Q_{lm} = \int d\bv{r} r^l n(\bv{r})Y^\star_{lm}(\hat{\bv{r}}),
\end{equation}
are the multipole moments of the density.
\end{subsection}


\begin{subsection}{Treatment of the exchange-correlation potential}
\label{sec:exch-corr}
All many-body effects are included within the
exchange-correlation potential, which in practice must be approximated. While many
sophisticated approximations to this potential have been developed~\cite{functionals}, the 
simplest is the adiabatic local density approximation (LDA) in the exchange-only limit 
(xLDA). In this case the exchange energy functional is given by 
\begin{equation}
E_x[n] = -\frac{3}{2} \left(\frac{3}{4\pi}\right)^{1/3}
	          \sum_{\sigma = \downarrow,\uparrow}\int d\bv{r}\,\, n_\sigma^{4/3}(\bv{r}, t),
\end{equation}
from which the exchange-correlation potential
\begin{equation}
V_{\mbox{\scriptsize xc}, \sigma}(\bv{r}, t) = - \left(\frac{6}{\pi}\right)^{1/3}
	          n_\sigma^{1/3}(\bv{r}, t),
\end{equation}
can be obtained. While this approximate functional is  easy to implement
it does suffer from the drawback of containing self-interaction errors. This self-interaction
means that the asymptotic form of the potential is exponential instead of Coulombic and many of the ground-state
properties of atoms and molecules can differ significantly from experimental values, as we shall see in 
Sec.~\ref{sec:resultsa2}. 
In spite of this problem, the LDA is one of the most widely used exchange-correlation functionals 
and will be used in this work in order to compare with previously published results.
\end{subsection}


\begin{subsection}{Treatment of the electron-ion potentials}
The Coulomb potential entering the external potential in the Kohn-Sham Hamiltonian, 
$V_{\mbox{\scriptsize ions}}(\bv{r}, \bv{R})$ as defined in Eq.~(\ref{eq:ion_pot}) with $\bv{r}_i$ replaced
by $\bv{r}$,
is singular whenever a finite-difference grid point coincides with one of the ions. Two approaches can be 
used to deal with these singularities depending on whether we wish to carry out all-electron calculations 
or if we only need to describe the response of valence electrons.
\begin{subsubsection}{All-electron calculations}
If we want to carry out simulations in which the response of all electrons must be included, for instance in 
simulating the response of molecules to XUV laser pulses, then we require an approach that accurately 
approximates the Coulomb potential associated with each ion. At the same time any singularities that 
arise must be removed. We follow the approach of Modine et al~\cite{modine:1997} who
calculate the Coulomb potential associated with each ion as the solution of a Poisson equation which approximates the volume charge density
associated with a point charge. In that case, a point charge carrying charge $Z_k$ located at the point $\bv{R}_k$ may
 be described by the volume charge distribution $Z_k\delta(\bv{r} - \bv{R}_k)$, where $\delta(\bv{x})$ is the
Dirac delta function. We approximate the Dirac delta function as
\begin{equation}
\delta(\bv{r} - \bv{R}_k) \approx A_k \exp\left[-\frac{(\bv{r} - \bv{\Theta}_k)^2}{\gamma_k^2}\right],
\end{equation}
for $\gamma_k > 0$. In this equation $A_k$ is a normalization factor and $\bv{\Theta}_k$ is a fictitious ion position. Both
these parameters are chosen to satisfy
\begin{equation}
\int d\bv{r} A_k \exp\left[-\frac{(\bv{r} - \bv{\Theta}_k)^2}{\gamma_k^2}\right] = 1,
\end{equation}
and
\begin{equation}
\int d\bv{r} \bv{r} A_k \exp\left[-\frac{(\bv{r} - \bv{\Theta}_k)^2}{\gamma_k^2}\right] = \bv{R}_k.	
\end{equation}
As in Sec.~\ref{sec:local_adapt}, $\bv{\Theta}_k \neq \bv{R}_k$ in general to allow for situations in 
which different locally adapted regions overlap.

Solution of the Poisson equation 
\begin{equation}
\nabla^2 V^D_k(\bv{r} - \bv{R}_k) = 4\pi Z_k A_k \exp\left[-\frac{(\bv{r} - \bv{\Theta}_k)^2}{\gamma_k^2}\right],
\label{eq:coulomb_dirac_ion}
\end{equation}
for ion $k$ then gives an approximation, $V^D_k(\bv{r} - \bv{R}_k)$, to the Coulomb 
potential so that
\begin{equation}
V_{\mbox{\scriptsize ions}}(\bv{r}, \bv{R}) \approx \sum_{k=1}^{N_n}V^D_k(\bv{r} - \bv{R}_k).
\label{eq:coulomb_dirac}
\end{equation}
The force acting of ion $k$ due to this potential will then be given by
\begin{eqnarray}
\bv{F}_k & = & -\int d\bv{r} n(\bv{r}) \gradi{k}V^D_k(\bv{x}_k) \nonumber \\
                                                       & = & \int d\bv{r} n(\bv{r}) \bv{\nabla} V^D_k(\bv{x}_k),
\end{eqnarray}
where $\bv{x}_k = \bv{r} - \bv{R}_k$.

\end{subsubsection}
\begin{subsubsection}{Pseudopotential calculations}
To describe the interaction of molecules with laser pulses operating at Ti:Sapphire wavelengths, we expect the laser
to couple predominantly to valence electrons. In that case all-electron calculations are not necessary and we 
can replace the Coulomb potentials with pseudopotentials. We implement norm-conserving Troullier-Martins
pseudopotentials~\cite{troullier:1991} in their fully-separable Kleinman-Bylander form~\cite{kleinman:1982}
\begin{widetext}
\begin{equation}
V_{\mbox{\scriptsize ions}}(\bv{r}, \bv{R}) = \sum_{k=1}^{N_n} 
\left[
V^{l_{\mbox{\scriptsize loc}}}_{k,\mbox{\scriptsize ps}} (\bv{x}_k)
+\mathop{\sum_{l}}_{l\ne l_{\mbox{\scriptsize loc}}}\sum_{m=-l}^{l} 
\frac{
\left|
\Delta V^{l}_{k,\mbox{\scriptsize ps}} (\bv{x}_k)
\chi^{k}_{lm}(\bv{x}_k)
\right>
\left<
\Delta V^l_{k,\mbox{\scriptsize ps}} (\bv{x}_k)
\chi^{k}_{lm}(\bv{x}_k)
\right|
}{
\left<{
\chi^{k}_{lm}(\bv{x}_k)
\left|
\Delta V^l_{k,\mbox{\scriptsize ps}} (\bv{x}_k)
\right|
\chi^{k}_{lm}(\bv{x}_k)
}\right>
}
\right]
\end{equation}
\end{widetext}
where $V^{l_{\mbox{\scriptsize loc}}}_{k,\mbox{\scriptsize ps}} (\bv{x}_k)$ defines the local
component of the pseudopotential of ion $k$, $\chi^{k}_{lm}(\bv{x}_k)$ denotes the pseudo-wavefunction for the partial
wave $\left|lm\right>$ and 
\begin{equation}
\Delta V^l_{k,\mbox{\scriptsize ps}} (\bv{x}_k) =
V^l_{k,\mbox{\scriptsize ps}} (\bv{x}_k) - 
V^{l_{\mbox{\scriptsize loc}}}_{k,\mbox{\scriptsize ps}} (\bv{x}_k),
\end{equation} 
define the non-local components of the pseudopotential.

The force acting on ion $k$ due to this non-local pseudopotential is given by~\cite{jing:1994}
\begin{eqnarray}
\bv{F}_k & = & \int d\bv{r} n(\bv{r}) \bv{\nabla}V^{l_{\mbox{\scriptsize loc}}}_{k,\mbox{\scriptsize ps}} (\bv{x}_k)\nonumber
\\
& + & \sum_{\sigma=\downarrow,\uparrow} \sum_{i=1}^{N_\sigma}\mathop{\sum_{l}}_{l\ne l_{\mbox{\scriptsize loc}}}\sum_{m=-l}^{l}
 \left(\frac{B_{lm}^{i\sigma k}}{A_{lm}^k}\right)\bv{\nabla}B_{lm}^{i\sigma k},
\end{eqnarray} 
where
\begin{equation}
A_{lm}^k = \left<{
\chi^{k}_{lm}(\bv{x}_k)
\left|
\Delta V^{\mbox{\scriptsize l}}_{k,\mbox{\scriptsize ps}} (\bv{x}_k)
\right|
\chi^{k}_{lm}(\bv{x}_k)
}\right>,
\end{equation}
and
\begin{equation}
B_{lm}^{i\sigma k} = 
\left<{\left.
\Delta V^{\mbox{\scriptsize l}}_{k,\mbox{\scriptsize ps}} (\bv{x}_k)
\chi^{k}_{lm}(\bv{x}_k)
\right|}\varphi_{i\sigma}(\bv{r}, t)\right>.
\end{equation}

In this paper all pseudopotentials were generated using the Atomic Pseudopotential Engine (APE)~\cite{oliveira:2008}.
\end{subsubsection} 
\end{subsection}


\begin{subsection}{Treatment of the laser-electron interaction}
Assuming that the dipole approximation is valid, we  define the vector potential of the laser pulse to be 
\begin{equation}
\bv{A}(t) = A(t)\hat{\bv{e}}
\end{equation}
where $\hat{\bv{e}}$ is the polarization direction and
\begin{equation}
A(t) = A_0 f(t) \cos(\omega_Lt + \phi),
\end{equation}
and where $\omega_L$ is the laser frequency, $\phi$ is the carrier-envelope phase (CEP) and $f(t)$ the pulse envelope 
given by
\begin{equation}
f(t) = 
\left\{
\begin{array}{ll}
\displaystyle\sin^2\left(\frac{\pi t}{T}\right) & 0\leq t \leq T\\
0 & \mbox{otherwise}
\end{array}
\right.,
\end{equation}
for a pulse of duration $T$.
For this choice of the vector potential, the electric field is given by $\bv{E}(t) = E(t)\hat{\bv{e}}$ where
\begin{equation}
E(t) = E_0f(t)\sin(\omega_L t + \phi) - \frac{E_0}{\omega_L} \frac{\partial f}{\partial t} \cos(\omega_Lt + \phi),
\end{equation}
and the peak electric field strength is related to the peak laser 
intensity, $I_0$, by
\begin{equation}
E_0 = \left(\frac{4\pi I_0}{c}\right)^{1/2}.
\end{equation}

The laser-electron interaction, $U_{\mbox{\scriptsize elec}}(\bv{r}, t)$, can be
represented in several ways. In this work a length gauge description of the
interaction is used, in which case
\begin{equation}
U_{\mbox{\scriptsize elec}}(\bv{r}, t) = U_L(\bv{r}, t) = \bv{r} \cdot \bv{E}(t).
\end{equation}
A velocity gauge description of the electron-field interaction can also be used 
but in previous grid calculations we have not observed any major differences in the results
for these two gauges~\cite{dundas:2004}.
\end{subsection}


\begin{subsection}{Wavefunction Splitting}
Ionizing wavepackets which reach the edge of the finite difference grid can be reflected from the 
boundary causing  spurious effects in both harmonic spectra and ionization 
rates. We eliminate these reflections by a splitting technique akin to an 
absorbing boundary which partitions the Kohn-Sham orbitals into two parts, one near, 
and the other far from the molecule where the Coulomb potential is 
negligible~\cite{dundas:2004,smyth:1998}. The splitting is implemented by a 
mask function, $M(\bv{r})$, which equals unity near the origin and goes asymptotically 
to zero very gradually. Using this mask function the Kohn-Sham orbital, $\varphi_{i\sigma}(\bv{r}, t)$, 
is split into two parts 
\begin{equation}
\varphi_{i\sigma}(\bv{r}, t) = M(\bv{r})\varphi_{i\sigma}(\bv{r}, t) + \left\{1 - M(\bv{r})\right\}\varphi_{i\sigma}(\bv{r}, t),
\end{equation}
The residual part, 
$\left\{1 - M(\bv{r})\right\}\varphi_{i\sigma}(\bv{r}, t)$, can be propagated independently in the limit in which the 
Coulomb potential is negligible over the region in which $M(\bv{r})<1$. The design and 
optimization of the mask function requires considerable care, and extensive 
numerical simulations must be performed to characterize the optimal 
shape~\cite{smyth:1998}. 

The mask function is written in the form
\begin{equation}
M(\bv{r}) = M_x(x) M_y(y) M_z(z),
\end{equation}
where $M_x(x)$ takes the form
\begin{equation}
M_x(x) = \left\{
\begin{array}{ll}
\displaystyle 1 & \displaystyle |x| \leq x_{\mbox{\scriptsize split}}\\[0.4cm]
\displaystyle \exp{\left[-{\left(
       \frac{x - x_{\mbox{\scriptsize start}}}
		               {\theta_x}\right)^2}\right]} & \displaystyle x_{\mbox{\scriptsize split}} < |x| \leq
			  x_{\mbox{\scriptsize max}}
\end{array}
\right..
\end{equation}
In this equation
\begin{equation}
\theta_x = \frac{x_{\mbox{\scriptsize max}} - x_{\mbox{\scriptsize split}}}{M_{\mbox{\scriptsize
edge}}},
\end{equation}
where $M_{\mbox{\scriptsize edge}} = M_x(x_{\mbox{\scriptsize max}})$ denotes the value of $M_x(x)$ that we
wish to impose at the boundary. Similar expressions are used to describe $M_y(y)$ and $M_z(z)$.

We note that the mask function does not have 
to equal zero at the boundary. The only requirement is that $\varphi_{i\sigma}(\bv{r}, t)$ smoothly 
approaches zero at the edges of the integration volume. 
\end{subsection}

\end{section}


\begin{section}{Results}
\label{sec:results}
In this section we will use our mixed quantum-classical approach to study the influence of multielectron 
effects in HHG in molecules. Before presenting these results, we will compare the 
performance of the eigensolvers described in Sec.~\ref{sec:initial}. In addition, the accuracy of ground-state 
properties will be compared with results obtained using a different method and with experiment. Our HHG 
results will consider N$_2$ and benzene interacting with linearly-polarized laser pulses. We will show how 
symmetries in the Kohn-Sham orbitals can lead to either enhancement and suppression of the harmonics as the orientation
between the laser and the molecule changes.


\begin{subsection}{Initial state generation: accuracy and efficiency}
\label{sec:resultsa}
The first stage of any simulation is obtaining the initial state of the system. In 
Sec.~\ref{sec:initial} three iterative approaches were described for calculating the ground-state
of a given system. In general we find that propagation in imaginary time only performs
efficiently when a small number of eigenpairs (less than five) are required. In light of this we
only compare the thick-restarted Lanczos method with the Chebyshev filtered subspace iteration method.
We then calculate ground state energies for a range of atoms and
diatomic molecules and show that accurate results can be obtained using a set of transferable 
grid adaptation parameters.

\begin{subsubsection}{Efficiency of the eigensolvers}
In order to compare the accuracy of the eigensolvers we consider the calculation of the ground-state of
N$_2$ using a globally adapted grid. The grid is set up so that the parameters controlling its extent 
and grid spacing are similar 
to those that will be used in the HHG simulations later. We
consider a grid that is globally adapted in the $x$ and $y$ directions using the transformation of 
Eq.~(\ref{eq:sinh_mesh}), while the $z$ coordinate is left unscaled. The grid spacings in each direction is the same ($\Delta \zeta^i = 0.4, i
= 1, 2, 3$) and the number of grid points used in each direction was $N_{\zeta^1} = N_{\zeta^2} = 57$
and $N_{\zeta^3} = 1001$. With appropriate scaling parameters in $x$ and $y$ the final grid extent was 
$-120\leq x \leq 120$, $-120\leq y \leq 120$ and $-200 \leq z \leq 200$. The calculation was
parallelized with the $z$ coordinate being distributed over 15 processors. Troullier-Martins norm-conserving 
pseudopotentials were used for the electron-ion interactions.

In table~\ref{tab:eigensolver} we present results for the time taken to obtain the ground state energy
for N$_2$: in all cases the convergence criteria used was that the least-squares norm of the difference
in density between self-consistent cycles was less than $10^{-7}$. It can clearly be seen that the 
CheFSI method outperforms the TRLan method by a factor of 7. For the
TRLan results, the calculation time reduces as we go to higher order in the Lanczos
method, however this comes at the cost of a larger memory requirement. For the CheFSI method 
the lower-order filter outperforms the higher-order filter. In this case the lower-order filter is
sufficient for the number of eigenpairs required. 

\begin{table}
\begin{center}
\begin{tabular*}{0.75\columnwidth}{@{\extracolsep{\fill}} ccc}
\hline\hline
Eigensolver & Order & \begin{tabular}{c}
Calculation time\\
(seconds)
\end{tabular}    \\\hline\hline
TRLan  & 18 & 8906 \\
       & 30 & 7638 \\\hline
CheFSI &  8 & 1051  \\
       & 15 & 1528  \\\hline\hline
\end{tabular*}
\end{center}
\caption{Comparison of the efficiency of the TRLan and CheFSI eigensolvers in calculating the ground
state of N$_2$. The time taken to obtain a converged, accurate ground state is given for various Lanczos
and Chebyshev filter orders. Other calculation parameters are detailed in the text.}
\label{tab:eigensolver}
\end{table}
\end{subsubsection}

\begin{subsubsection}{Accuracy of the energy eigenstates using locally adaptive grids}
\label{sec:resultsa2}

\begin{table*}
\begin{tabular*}{0.7\textwidth}{@{\extracolsep{\fill}} crrrrr} \hline\hline
\multicolumn{1}{c}{Atom} & 
\multicolumn{3}{c}{Ground State Energy} &
\multicolumn{2}{c}{Ionization potential}\\
   & \multicolumn{1}{r}{Present} 
   & \multicolumn{1}{r}{Grabo~\protect\cite{grabo:2000}}
   & \multicolumn{1}{r}{Experimental~\protect\cite{grabo:2000}}
   & \multicolumn{1}{r}{Present}
   & \multicolumn{1}{r}{Experimental~\protect\cite{grabo:2000}}
 \\  \cline{2-4}
\hline\hline
H  & $-$0.4568   & $-$0.4571  & $-$0.5000  & 0.4568 & 0.5000 \\
He & $-$2.7191   & $-$2.7236  & $-$2.9037  & 0.8442 & 0.9037 \\
Li & $-$7.1741   & $-$7.1934  & $-$7.4781  & 0.1767 & 0.1982 \\
Be & $-$14.1834  & $-$14.2233 & $-$14.6674 & 0.2835 & 0.3426 \\
B  & $-$24.0566  & $-$23.7791 & $-$24.6539 & 0.2775 & 0.3050 \\
C  & $-$37.0977  & $-$37.1119 & $-$37.8450 & 0.3897 & 0.4140 \\
N  & $-$53.6854	 & $-$53.7093 & $-$54.5893 & 0.4890 & 0.5348 \\
\hline\hline
\end{tabular*}
\caption{xLDA all-electron atomic ground state energies and ionization potentials calculated 
using a locally adapted finite difference grid in 3D. The results of the present work are 
compared with experiment and with the theoretical calculations of Grabo et al~\cite{grabo:2000}. 
The current results agree
well with the xLDA results of Grabo et al. Both ground state energies and ionization potentials are
generally underestimated using xLDA. The ionization potentials, calculated as vertical ionization potentials,
show much better agreement with experiment than the ground state energies. The grid adaptation and 
Coulomb potential parameters used were chosen to reproduce the correct hydrogenic energy levels 
of nitrogen: these parameters are used for all other species considered. Other calculation parameters 
are detailed in the text.}
\label{tab:energies_atoms}
\end{table*}
To show the accuracy of our approach in the calculation of the initial state when 
using locally adaptive grids, we have compared the static properties of a range of atoms and 
molecules with those obtained by Grabo et al~\cite{grabo:2000} and with experiment. 
Table~\ref{tab:energies_atoms} presents results for atoms, table~\ref{tab:energies_molecules} presents 
results for diatomic molecules and table~\ref{tab:energies_benzene} presents results for benzene. For all
calculations, we consider a grid that is locally adapted  using the transformation given in
Eq.~(\ref{eq:adapt_tranform}). A 5-point finite difference rule was used and the grid spacings 
in each direction were kept the same ($\Delta \zeta^i = 0.2, i
= 1, 2, 3$). The number of grid points used was $N_{\zeta^1} = N_{\zeta^2} = 
N_{\zeta^3} = 201$ so that the  grid extent was 
$-20\leq x \leq 20$, $-20\leq y \leq 20$ and $-20 \leq z \leq 20$. 

In these calculations all electrons 
are considered and the Coulomb potential associated with each ion calculated  using
Eq.~(\ref{eq:coulomb_dirac_ion}). Appropriate grid adaptation and Coulomb potential parameters were chosen as
follows. The atom with the largest charge is nitrogen. The parameters for this atom were chosen by 
tuning them to obtain the correct hydrogenic energy levels for nitrogen (by switching off the Hartree and exchange correlation 
potentials). The resulting parameters were then used for all other atoms and molecules. The results show that 
not only are accurate energies obtained (when compared with other xLDA calculations), but the adapted grid is transferrable to other atomic and molecular systems 
having smaller nuclear charge. However, it is clear that the properties calculated using the xLDA
exchange-correlation functional differ significantly from experimental values. This is due to the
limitations of the LDA as outlined in Sec.~\ref{sec:exch-corr}. Even so, the ionization potentials obtained from these
calculations are in fairly good agreement with experiment. These have been calculated as
vertical ionization potentials, rather than estimating them from the energy of the highest occupied
molecular orbital (HOMO).

\begin{table*}
\begin{tabular*}{0.85\textwidth}{@{\extracolsep{\fill}} crrrrrr} \hline\hline
\multicolumn{1}{c}{Molecule} & 
\multicolumn{2}{c}{Bond Length, $R_e$} &
\multicolumn{2}{c}{Dissociation Energy, $D_e$} &
\multicolumn{2}{c}{Ionization Potential}\\
   & \multicolumn{1}{r}{Present} 
   & \multicolumn{1}{r}{Experimental~\protect\cite{grabo:2000}} 
   & \multicolumn{1}{r}{Present} 
   & \multicolumn{1}{r}{Experimental~\protect\cite{grabo:2000}}
   & \multicolumn{1}{r}{Present}
   & \multicolumn{1}{r}{Experimental~\protect\cite{grabo:2000}}
 \\ 
\hline\hline
N$_2$   & 2.071 & 2.074  & 0.381 & 0.364 & 0.517  & 0.573  \\
LiH     & 3.120 & 3.015  & 0.080 & 0.092 & 0.267  & 0.283  \\
BH      & 2.414 & 2.329  & 0.148 & 0.135 & 0.308  & 0.359  \\
\hline\hline
\end{tabular*}
\caption{xLDA all-electron equilibrium molecular bond lengths, dissociation energies and ionization potentials
using a locally adapted finite difference grid in 3D. The results of the present work are compared 
with experimental results. As in the atomic case, we see that ionization potentials are underestimated using xLDA.
The grid adaptation and Coulomb potential parameters used were chosen to reproduce the correct hydrogenic energy levels 
in the nitrogen atom: these parameters are used for all other species considered. Other calculation 
parameters are detailed in the text.}
\label{tab:energies_molecules}
\end{table*}

\begin{table*}
\begin{tabular*}{\textwidth}{@{\extracolsep{\fill}} rrrrrrrr} \hline\hline
\multicolumn{4}{c}{Equilibrium geometry} &
\multicolumn{2}{c}{Atomization energy} &
\multicolumn{2}{c}{Ionization Potential}\\
\multicolumn{2}{c}{C$-$C bond length} &
\multicolumn{2}{c}{C$-$H bond length} &
\multicolumn{4}{c}{} \\
   \multicolumn{1}{r}{Present} & \multicolumn{1}{r}{Experimental~\protect\cite{meijer:1996}} 
 & \multicolumn{1}{r}{Present} & \multicolumn{1}{r}{Experimental~\protect\cite{meijer:1996}}
 & \multicolumn{1}{r}{Present} & \multicolumn{1}{r}{Experimental~\protect\cite{parthiban:2001}}
 & \multicolumn{1}{r}{Present} & \multicolumn{1}{r}{Experimental~\protect\cite{sharifi:2007}}
 \\ 
\hline\hline
2.645 & 2.644 & 2.091 & 2.081 & 3.283 & 2.081 & 0.292 & 0.340 \\
\hline\hline
\end{tabular*}
\caption{xLDA all-electron static properties of benzene. The equilibrium C--C and C--H bond lengths,
atomization energy and ionization potential are compared with experiment. We see that the atomization energy is
greatly overestimated, in common with other LDA calculations~\protect\cite{lehtovaara:2009,son:2011}. The grid adaptation and 
Coulomb potential parameters used were chosen to reproduce the correct hydrogenic energy levels 
in the nitrogen atom. Other calculation parameters are detailed in the text.}
\label{tab:energies_benzene}
\end{table*}
\end{subsubsection}
\end{subsection}


\begin{subsection}{Multielectron and orientation effects in molecular HHG}
We now consider HHG in molecules irradiated by intense, short-duration laser pulses. 
According to classical electromagnetism the emission of secondary radiation arises from accelerating 
dipole moments induced by the laser pulse. The emission of high-order harmonics is generally understood in terms of the
three-step model of Corkum~\cite{corkum:1993} and Kulander et al~\cite{kulander:1992} in which electrons initially tunnel through the
field-modified Coulomb potential barrier, propagate in the laser field and eventually recombine with the ion.
HHG spectra are calculated readily within 
TDDFT since only a simple functional of the electron density is involved. The spectral density
is calculated from the Fourier transform of the dipole acceleration~\cite{burnett:1992}
\begin{equation}
S(\omega) = \left|\int dt\,\, e^{{\rm i}\omega t}
\hat{\bv{e}}\cdot\ddot{\bv{d}}(t)\right|^2.
\end{equation}
The dipole acceleration, $\ddot{\bv{d}}(t)$, can be calculated via Ehrenfest's 
theorem as
\begin{equation}
\ddot{\bv{d}}(t) = -\sum_{\sigma=\downarrow,\uparrow}
\int n_\sigma(\bv{r}, t)\bv{\nabla}
V_{\mbox{\scriptsize eff}, \sigma}(\bv{r}, \bv{R}, t) d\bv{r},
\label{eq:dipole-acceleration}
\end{equation}
where $V_{\mbox{\scriptsize eff}, \sigma}(\bv{r}, \bv{R}, t)$ is the effective potential 
given by Eq.~(\ref{eq:tdks_veff}). When using the LDA approximation, described in Sec.~\ref{sec:exch-corr}, 
it must be noted that the incorrect asymptotic behaviour of the exchange-correlation potential means that
recollisions are not described with quantitative accuracy~\cite{bandrauk:2010}. However, the main qualitative 
features of HHG spectra are reproduced, as can be seen when LDA results are compared with asymptotically correct
functionals~\cite{bandrauk:2010}.  

\begin{figure*}
\centerline{\hfill\includegraphics[width=4cm]{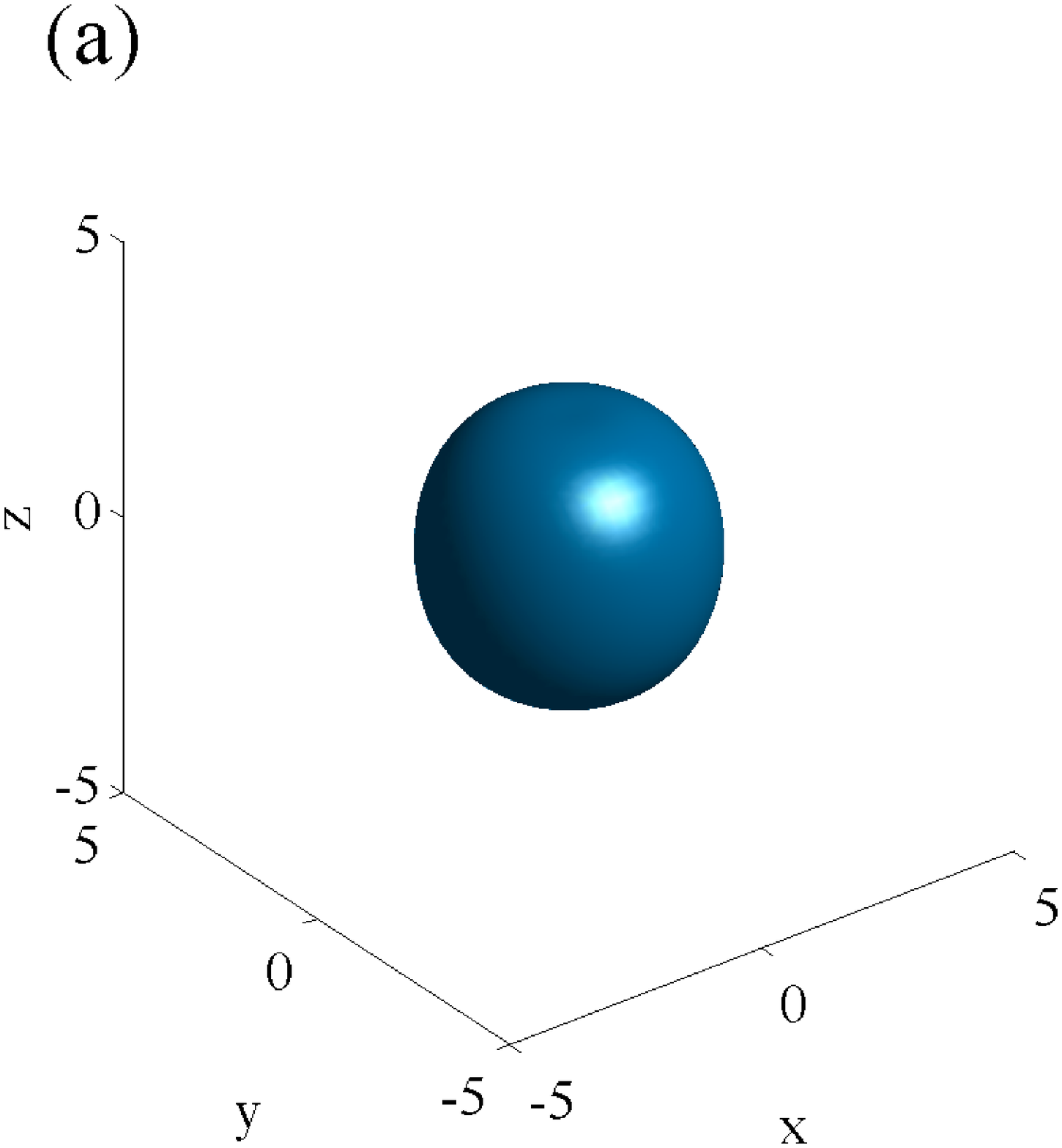}
            \hfill\includegraphics[width=4cm]{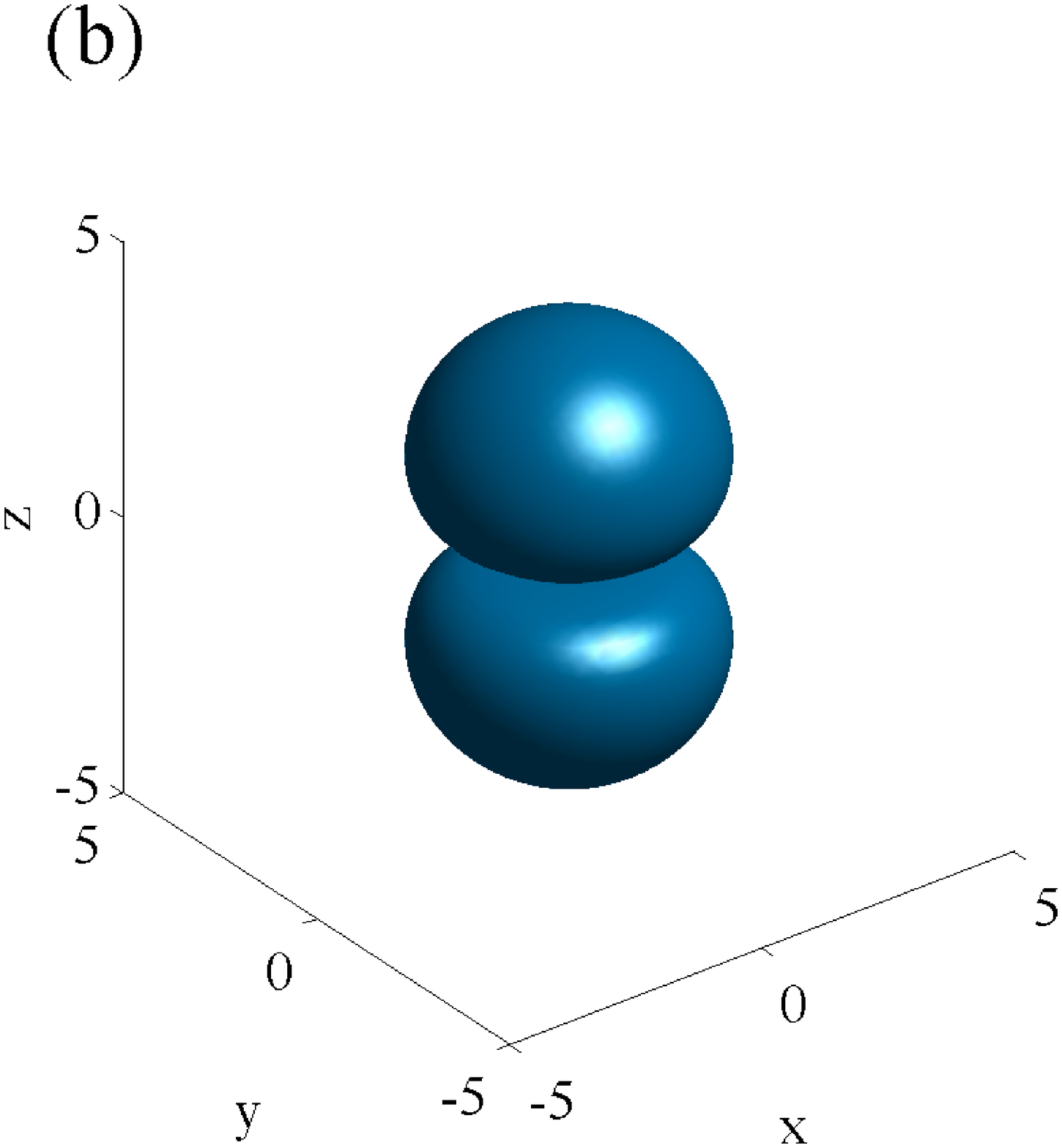}
	    \hfill\includegraphics[width=4cm]{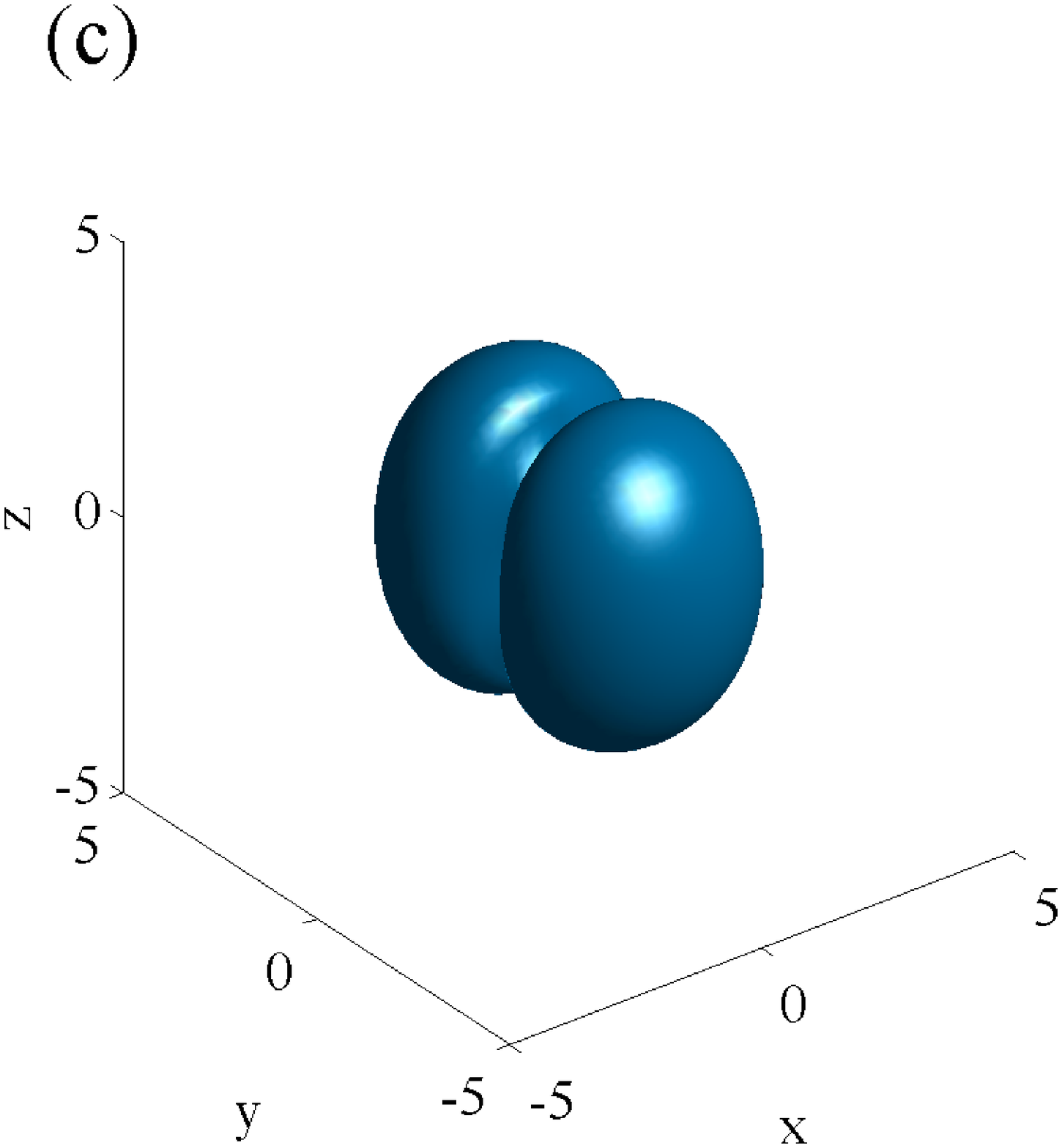}\hfill}
\centerline{\hfill\includegraphics[width=4cm]{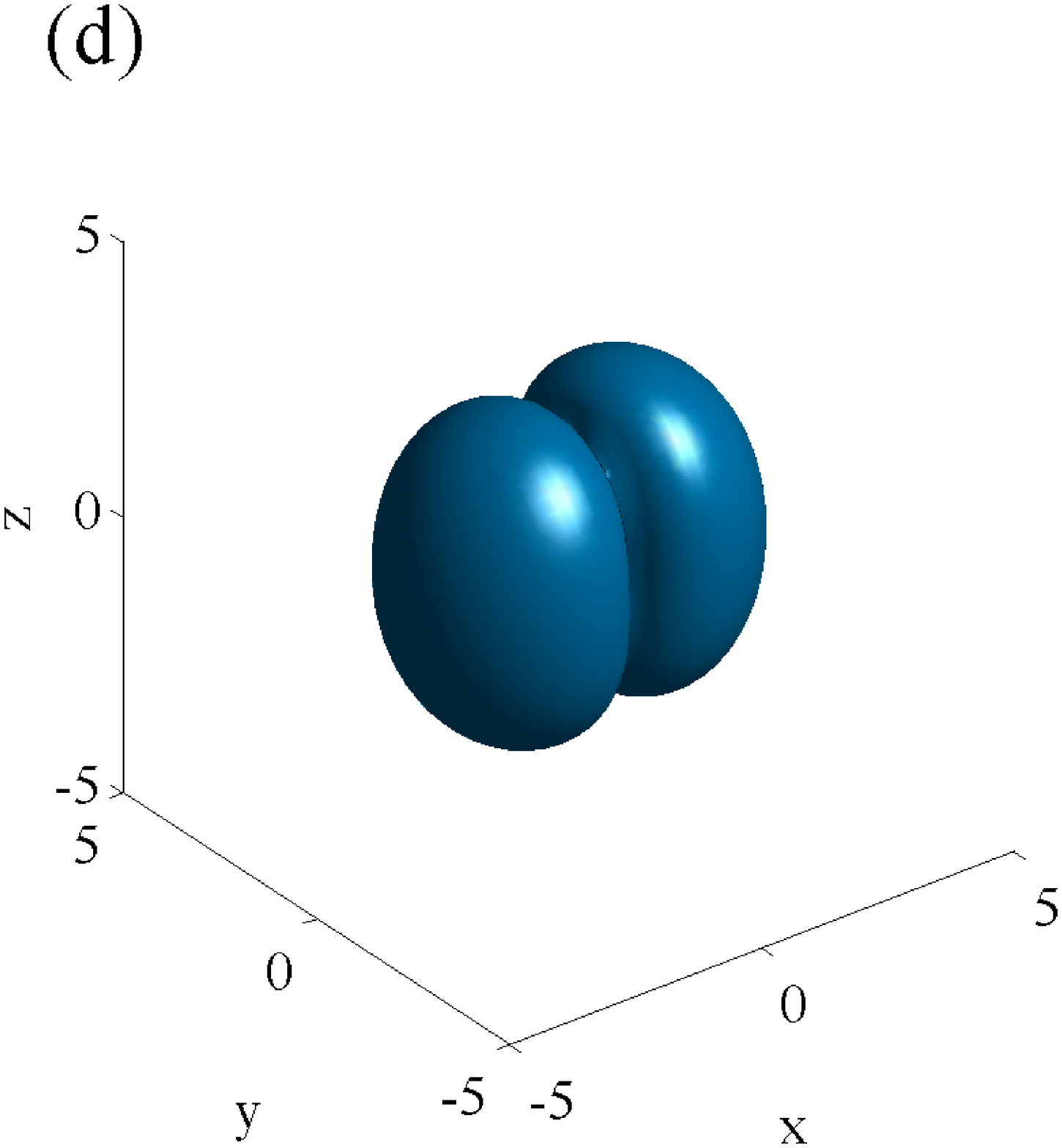}
            \hfill\includegraphics[width=4cm]{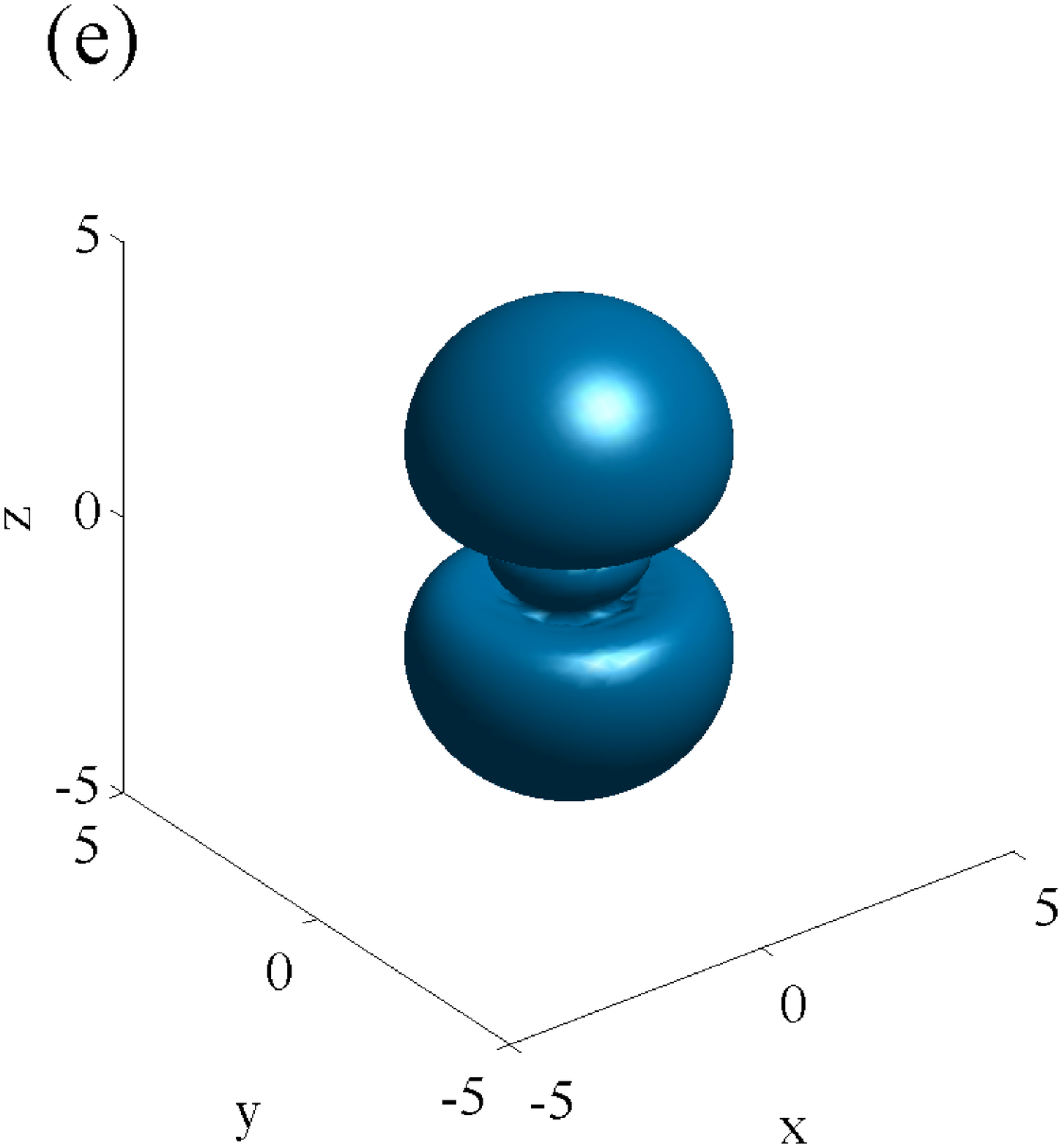}\hfill}
\caption{(Color online) Isosurface plots of the valence Kohn-Sham orbital densities in N$_2$ 
considered in our calculations. The occupation of each Kohn-Sham orbital is 2. The ground state energy
configuration of N$_2$ is $1\sigma_g^21\sigma_u^22\sigma_g^22\sigma_u^21\pi_u^43\sigma_g^2$ and so
each orbital is labeled as follows: (a)$=2\sigma_g$, (b)$=2\sigma_u$,
(c)$=1\pi_u$, (d)$=1\pi_u$ and (e)$=3\sigma_g$. In these plots, the molecular axis is
aligned  along the $z$-axis and for the HHG results the laser polarization will either be along the $x$- or
the $z$-axis.}
\label{fig:orbitals_n2}
\end{figure*}

Using our approach we will now consider HHG when a linearly polarized laser pulse interacts with  N$_2$ and 
benzene. In all simulations the following parameters are used. We fix the laser polarization direction along either the
$x$-axis or the $z$-axis. When the laser is aligned along the $z$-axis, we use a grid that is unadapted along the $z$-axis 
while the $x$- and $y$-axes are globally adapted using the transformation of Eq.~(\ref{eq:sinh_mesh}). A 5-point finite 
difference rule will be used with equal grid spacings in each coordinate ($\Delta \zeta^i = 0.4, i = 1, 2, 3$). The number 
of grid points used for each coordinate is $N_{\zeta^1} = N_{\zeta^1} = 61$ and $ N_{\zeta^3} = 999$. With appropriate 
scaling parameters the final grid extent is $-120\leq x\leq 120$, $-120\leq y\leq 120$ and $-199.6\leq z\leq 199.6$. 
In the case where the laser is aligned along the $x$-axis, the grid parameters for the $x$- and $z$-axes are interchanged:
we now globally adapt the grid along the $z$-axis while the grid remains unadapted along the $x$-axis and the extent of the
grid in $x$ and $z$ becomes  $-199.6\leq x\leq 199.6$ and $-120\leq z\leq 120$.
Troullier-Martins norm-conserving 
pseudopotentials were used for the electron-ion interactions and time propagation was carried out using a 6th-order Arnoldi propagator
with a time step $\Delta t_e = 0.04$.

\begin{subsubsection}{Multielectron and orientation effects in HHG in N$_2$}

Our calculations for N$_2$ will consider its interaction with a 10-cycle linearly-polarized Ti-Sapphire ($\lambda =$ 780nm) 
laser pulse having a peak intensity of \intensity{4.0}{14}. The internuclear spacing is set to 2.07$a_0$ and the molecule is 
aligned along the $z$-axis. For the two orientations of the laser considered, the polarization direction is either 
parallel (along the $z$-axis) or perpendicular (along the $x$-axis) to the molecular axis. All ions are kept fixed 
throughout the simulation. We consider the 5 valence Kohn-Sham orbitals in our simulations which are plotted in
Fig.~\ref{fig:orbitals_n2}. Each orbital is labeled (a)--(e) as detailed in the figure. We can see that 
orbitals (c) and (d) will respond in a similar fashion to the field whenever the laser polarization is 
parallel to the molecular axis whereas they will respond differently whenever the laser polarisation is perpendicular to the 
molecular axis. It must be stressed at this point that the Kohn-Sham orbitals do not have any direct connection to the 
actual molecular orbitals. However, studying the evolution of the Kohn-Sham orbitals allows us to 
obtain information about the importance of orbital symmetries in the response and is widely used in TDDFT studies. 

Fig.~\ref{fig:hhg_n2} presents the HHG spectra.
The cut-off in the plateau region of the spectrum is given by 
\begin{equation}
E_{\mbox{\scriptsize cut-off}} = 1.32I_p + 3.17U_p, 
\label{eq:cut-off-law}
\end{equation}
where $I_p$ is the ionization potential of the molecule and $U_p$ is the laser ponderomotive 
energy~\cite{lewenstein:1994}. Thus, for the laser parameters considered the cut-off should occur at 
harmonic 57, and indeed we see that our calculated cut-off agrees quite well with this value. 
For harmonic orders less than 21 the spectral density is greatest whenever the laser polarization 
is parallel to the molecular axis, while for higher harmonics the converse is true. Fig.~\ref{fig:pop_n2} presents 
the orbital populations during the interaction with the laser pulse perpendicular to the molecular axis. It is 
clear that the two forms of the $1\pi_u$ molecular orbital (the HOMO-1 orbital) -- (c) and (d) in 
Figure~\ref{fig:orbitals_n2} -- respond differently to the field and that the $1\pi_u$ (d) orbital does indeed 
respond more than the HOMO. This is in contrast to the parallel orientation where both $1\pi_u$ orbitals respond 
identically and where the HOMO responds predominantly. This type of behaviour is similar to that observed in 
ionization studies of OCS and CS$_2$~\cite{bandrauk:2011}. However, in previous DFT studies of N$_2$ by 
Petretti et al~\cite{petretti:2010}, the $1\pi_u$ orbital was observed to show a lower response  than the HOMO in
perpendicular alignment. These previous calculations were performed at a lower laser
intensity and it is unclear if the two forms of the $1\pi_u$ orbitals were included explicitly in their calculations.

We note that in a previous experiment, McFarland et al~\cite{mcfarland:2008} showed that in the
perpendicular alignment the harmonic signal was enhanced, relative to the  parallel case, around the 
cut-off in the plateau. This enhancement was attributed to the contribution from electrons in the more 
tightly bound $1\pi_u$ HOMO-1 orbital compared with electrons in the 
$3\sigma_g$ HOMO. The influence of the HOMO-1 orbital in HHG in the perpendicular orientation has also been studied
in a more recent experiment~\cite{jin:2012}. However, the results of McFarland et al disagree with earlier
experiments in which the harmonic signal was greater in the parallel orientation than in the perpendicular 
orientation~\cite{mairesse:2008,haessler:2010,rupenyan:2012}. These earlier experiments 
were carried out at lower laser intensities and mainly considered plateau harmonics. It has been suggested 
that the results of McFarland et al may be due to propagation effects~\cite{sickmiller:2009}: these effects have not 
been considered in the present work. Additionally, in a range of 
calculations based on quantitative rescattering theory it has been found that the contribution of the 
HOMO-1 orbital becomes important in the perpendicular alignment at high laser intensities when considering both the
single atom response~\cite{le:2009} and when including propagation effects~\cite{jin:2012}. 

\begin{figure}
\centerline{\includegraphics[width=8cm]{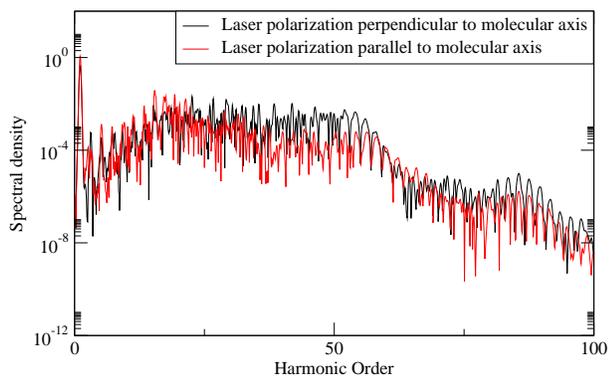}}
\caption{(Color online) HHG spectra for N$_2$.
The molecule interacts with a 10-cycle linearly polarized Ti-Sapphire ($\lambda =$ 780nm) laser 
pulse having a peak intensity of \intensity{4.0}{14}. The molecular axis is aligned along the $z$-axis. 
The black line shows the spectrum when the
laser polarization lies perpendicular to the molecular axis while the red line denotes the spectrum when the
laser polarization lies parallel to the molecular axis.}
\label{fig:hhg_n2}
\end{figure}

\begin{figure}
\centerline{\includegraphics[width=8cm]{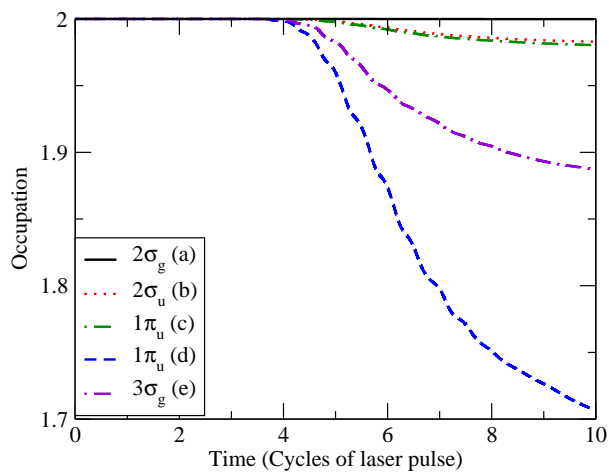}}
\caption{(Color online) Occupation of the valence Kohn-Sham orbitals of N$_2$ during interaction with a 10-cycle linearly 
polarized Ti-Sapphire ($\lambda =$ 780nm) laser 
pulse having a peak intensity of \intensity{4.0}{14}. The laser polarization direction is perpendicular to the
molecular axis. We see that the more tightly bound $1\pi_u$ (d) orbital responds more than the $3\sigma_g$ 
HOMO. The change in the occupation of the $2\sigma_g$ orbital during the interaction with the pulse is too small to
show up on this scale.}
\label{fig:pop_n2}
\end{figure}

\end{subsubsection}

\begin{subsubsection}{Multielectron and orientation effects in HHG in benzene}
While most experimental studies of
HHG in benzene consider the interaction with a linearly polarized laser pulse~\cite{hay:2000}, theoretical 
studies generally study the response to circularly polarized light~\cite{zdanska:2003,baer:2003,ceccherini:2001}. In our
simulations we will consider HHG using a 10-cycle linearly polarized Ti-Sapphire ($\lambda =$ 780nm) laser 
pulse having a peak intensity of \intensity{4.0}{14}. The benzene molecule lies 
in the $x-y$ plane with the atom positions as shown in Fig.~\ref{fig:adapt}. For the two laser alignments the polarization direction
is either parallel (along the $x$-axis) or perpendicular (along the $z$-axis) to the molecular plane. All ions are kept fixed throughout the 
simulation. Fig.~\ref{fig:hhg_benzene} presents harmonic spectra. For the laser parameters used, the cut-off region for the
plateau, using Eq.~(\ref{eq:cut-off-law}), should occur around harmonic 53. We see that our results are in good agreement.
The intensity of those harmonics of order less than 21 are comparable while for higher-order harmonics the harmonic
intensity is greatest in the perpendicular orientation. The reason for the suppression in the parallel orientation can be understood 
by considering the symmetry of the HOMO orbital. In benzene the HOMO is doubly degenerate: isosurface plots of these two orbitals are presented in
Fig.~\ref{fig:orbitals_benzene} and are referred to as HOMO~(a) and HOMO~(b). 
Fig.~\ref{fig:pop_benzene} presents the populations for the Kohn-Sham orbitals in the parallel orientation during the interaction with the pulse: 
for clarity we have only labeled the HOMO~(a) and HOMO~(b) orbitals. 
We see that the response of the HOMO~(b) orbital 
is suppressed, relative to the HOMO~(a). In the perpendicular alignment both HOMO orbitals respond in the same way to the field and the response is similar to that of the 
HOMO~(a) in Fig.~\ref{fig:pop_benzene}. This suggests that the reduction in the intensity of plateau harmonics 
near the cut-off  in the parallel alignment is due to the symmetry-induced suppression of the HOMO~(b) orbital.
\end{subsubsection}
\begin{figure}
\centerline{\includegraphics[width=8cm]{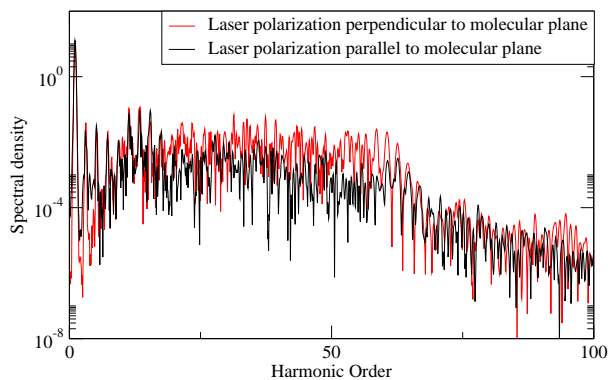}}
\caption{(Color online) HHG spectra for benzene.
The molecule interacts with a 10-cycle linearly polarized Ti-Sapphire ($\lambda =$ 780nm) laser 
pulse having a peak intensity of \intensity{4.0}{14}. The molecule lies in the $x-y$ plane as shown if Fig.~\protect\ref{fig:adapt}. 
The red line shows the 
spectrum when the laser-polarization lies perpendicular to the molecular plane while the black line denotes the 
spectrum when the laser-polarization lies parallel to the plane of the molecule.}
\label{fig:hhg_benzene}
\end{figure}

\begin{figure}
\centerline{\includegraphics[width=4cm]{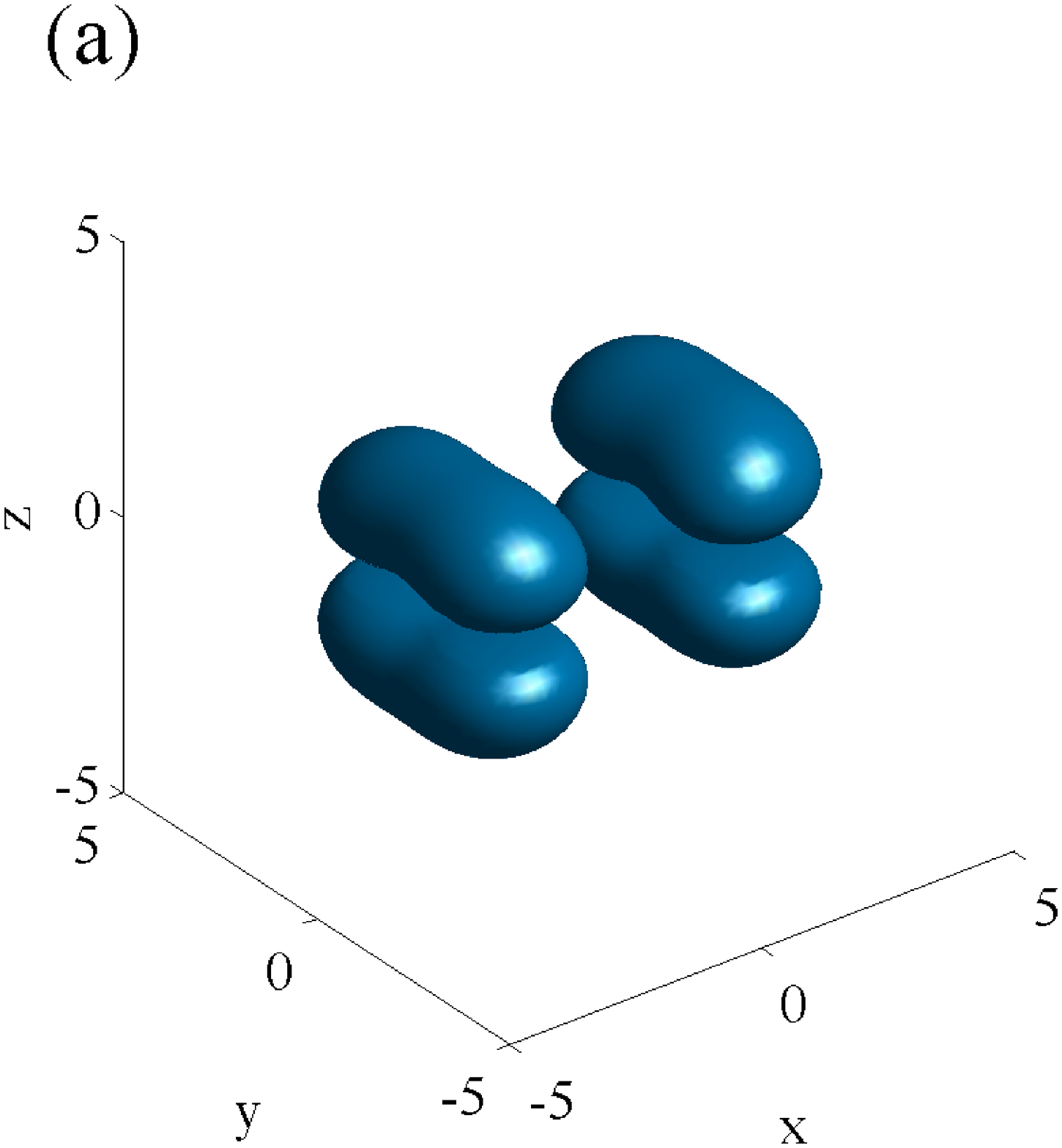}}
\centerline{\includegraphics[width=4cm]{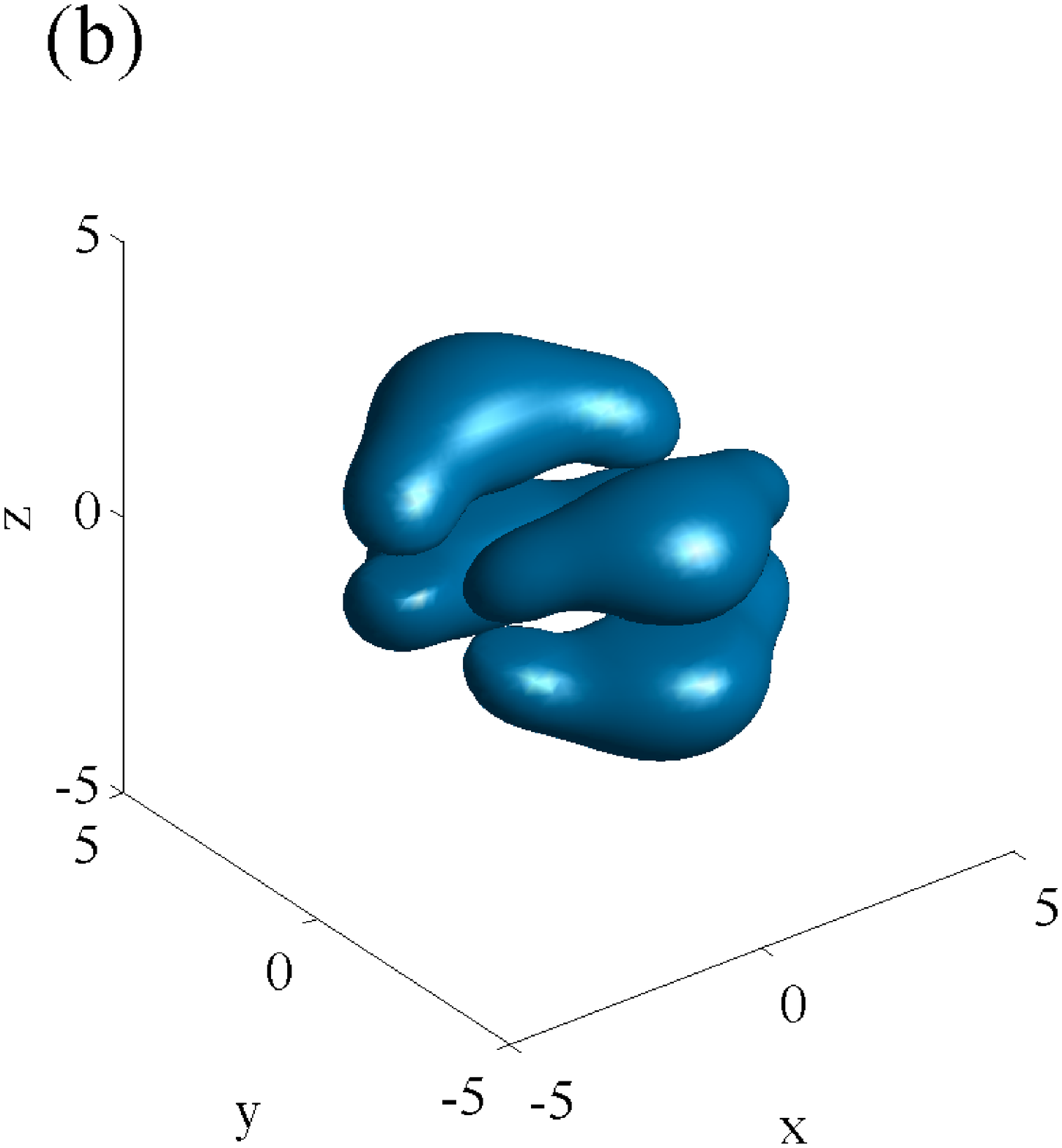}}
\caption{(Color online) Isosurface plots of the two degenerate HOMO Kohn-Sham orbital densities for 
benzene in our calculations. The molecule lies in the $x-y$ plane as denoted in Fig.~\protect\ref{fig:adapt}. 
The occupation of each Kohn-Sham orbital is 2. We label these orbitals as HOMO~(a) and HOMO~(b). 
In these plots, the molecule lies in the $x-y$ plane and for the HHG results the laser polarization will either be along the $x$- or
the $z$-axis.}
\label{fig:orbitals_benzene}
\end{figure}

\begin{figure}
\centerline{\includegraphics[width=8cm]{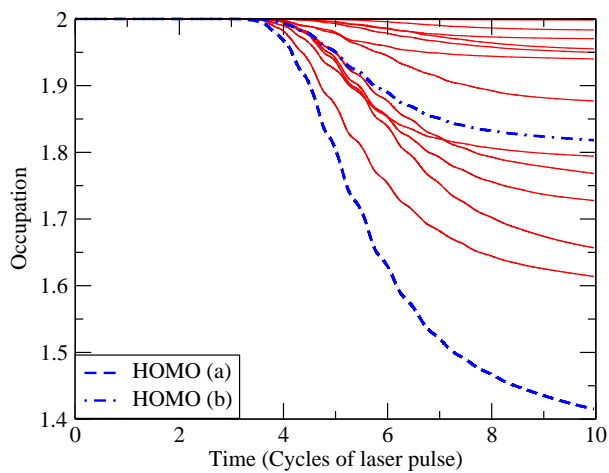}}
\caption{(Color online) Occupation of the valence Kohn-Sham orbitals of benzene during interaction with a 10-cycle 
linearly polarized Ti-Sapphire ($\lambda =$ 780nm) laser pulse having a peak intensity of \intensity{4.0}{14}. 
The molecule lies in the $x-y$ plane and the laser-polarization lies along the $x$-axis (parallel to the molecular
plane). For clarity only the HOMO~(a) and
HOMO~(b) orbital populations are labeled, with the response of all other orbitals shown in red. We see that the
response of the HOMO~(b) orbital is suppressed with respect to
the HOMO~(a) orbital.
}
\label{fig:pop_benzene}
\end{figure}

\end{subsection}

\end{section}


\begin{section}{Conclusions}
\label{sec:conclusions}
In this paper we have presented a NAQMD 
approach for treating laser-molecule interactions. The approach is of sufficient
generality that the response of a wide range of molecules can be studied: in
particular both all-electron and pseudopotential calculations can be performed.
Using this approach we were able to study
the role of multielectron effects in HHG in N$_2$ and benzene and show how
the symmetry properties of the Kohn-Sham orbitals play an important role in the 
observed spectra. To the best of our knowledge, the results for benzene represent the 
first TDDFT calculations of HHG in benzene using linearly polarized 
laser pulses and suggest a suppression of the harmonic signal when the laser polarization direction
is aligned perpendicular to the molecular plane. 

The results presented here consider only fixed nuclei and exchange-correlation
is only treated at the level of the LDA. However, our results 
are able to capture important aspects of the molecular response to the laser pulse.
Subsequent studies will consider dynamical ion calculations. This will allow charge transfer
across molecules during interaction with the laser pulse to be studied~\cite{remacle:2006,lunnermann:2008} 
as well as how this transfer influences dissociation of the molecule. In addition,
asymptotically-corrected exchange-correlation potentials~\cite{vanleeuwen:1994} will be employed
and this will be used to assess the accuracy of the excited states properties in our calculations 
and their importance in HHG~\cite{mairesse:2010}.

Another important extension to the approach is the incorporation of
transport boundary conditions that will allow current-flow through molecular
electronic devices to be considered. Several schemes for achieving this are
being studied~\cite{mceniry:2007,stefanucci:2008,cheng:2006}.
\end{section}

\end{document}